\documentclass[sigconf]{acmart}
\usepackage{graphicx}
\usepackage{subcaption}
\usepackage{algorithm}
\usepackage{algpseudocode}
\usepackage{enumitem}
\usepackage{multirow}
\usepackage{tabularx}
\usepackage{threeparttable}
\usepackage{color, xspace}
\usepackage{appendix}
\usepackage{balance}
\usepackage{amsmath}
\usepackage{fontawesome5}
\usepackage{amsmath}
\AtBeginDocument{%
  }

\setcopyright{acmlicensed}
\copyrightyear{2018}
\acmYear{2018}
\acmDOI{XXXXXXX.XXXXXXX}
\acmConference[Conference acronym 'XX]{Make sure to enter the correct
  conference title from your rights confirmation email}{June 03--05,
  2018}{Woodstock, NY}

\acmISBN{978-1-4503-XXXX-X/2018/06}

\begin{document}

\title{DUET: Dual Model Co-Training for Entire Space CTR Prediction}

\author{Yutian Xiao}
\affiliation{%
  \institution{Kuaishou Technology Co., Ltd.}
  \streetaddress{}
  \city{Beijing}
  \state{}
  \country{China}
}
\email{xiaoyutian@kuaishou.com}

\author{Meng Yuan}
\affiliation{%
  \institution{Kuaishou Technology Co., Ltd.}
  \streetaddress{}
  \city{Beijing}
  \state{}
  \country{China}
}
\email{yuanmeng05@kuaishou.com}

\author{Fuzhen Zhuang}
\affiliation{%
  \institution{Independent Researcher}
  \streetaddress{}
  \city{Beijing}
  \state{}
  \country{China}
}
\email{zfz20081983@gmail.com}

\author{Wei Chen}
\affiliation{%
  \institution{Independent Researcher}
  \streetaddress{}
  \city{Beijing}
  \state{}
  \country{China}
}
\email{cwei_01@163.com}

\author{Shukuan Wang}
\affiliation{%
  \institution{Kuaishou Technology Co., Ltd.}
  \streetaddress{}
  \city{Beijing}
  \state{}
  \country{China}
}
\email{wangshukuan@kuaishou.com}

\author{Shanqi Liu}
\affiliation{%
  \institution{Kuaishou Technology Co., Ltd.}
  \streetaddress{}
  \city{Beijing}
  \state{}
  \country{China}
}
\email{liushanqi@kuaishou.com}

\author{Chao Feng}
\authornote{Corresponding authors.}
\affiliation{%
  \institution{Kuaishou Technology Co., Ltd.}
  \streetaddress{}
  \city{Beijing}
  \state{}
  \country{China}
}
\email{fengchao08@kuaishou.com}

\author{Wenhui Yu}
\affiliation{%
  \institution{Independent Researcher}
  \streetaddress{}
  \city{Beijing}
  \state{}
  \country{China}
}
\email{naywh@qq.com}

\author{Xiang Li}
\affiliation{%
  \institution{Kuaishou Technology Co., Ltd.}
  \streetaddress{}
  \city{Beijing}
  \state{}
  \country{China}
}
\email{lixiang44@kuaishou.com}

\author{Lantao Hu}
\affiliation{%
  \institution{Kuaishou Technology Co., Ltd.}
  \streetaddress{}
  \city{Beijing}
  \state{}
  \country{China}
}
\email{hulantao@kuaishou.com}

\author{Han Li}
\affiliation{%
  \institution{Kuaishou Technology Co., Ltd.}
  \streetaddress{}
  \city{Beijing}
  \state{}
  \country{China}
}
\email{lihan08@kuaishou.com}

\author{Zhao Zhang}
\affiliation{%
  \institution{Independent Researcher}
  \streetaddress{}
  \city{Beijing}
  \state{}
  \country{China}
}
\email{zhangzhao.cs.ai@gmail.com}

\renewcommand{\shortauthors}{Trovato et al.}

\begin{abstract}
The pre-ranking stage plays a pivotal role in large-scale recommender systems but faces an intrinsic trade-off between model expressiveness and computational efficiency. Owing to the massive candidate pool and strict latency constraints, industry systems often rely on lightweight two-tower architectures, which are computationally efficient yet limited in estimation capability. As a result, they struggle to capture the complex synergistic and suppressive relationships among candidate items, which are essential for producing contextually coherent and diverse recommendation lists. Moreover, this simplicity further amplifies the Sample Selection Bias (SSB) problem, as coarse-grained models trained on biased exposure data must generalize to a much larger candidate space with distinct distributions.

To address these issues, we propose \textbf{DUET} (\textbf{DU}al Model Co-Training for \textbf{E}ntire Space C\textbf{T}R Prediction), a set-wise pre-ranking framework that achieves expressive modeling under tight computational budgets. Instead of scoring items independently, DUET performs set-level prediction over the entire candidate subset in a single forward pass, enabling information-aware interactions among candidates while amortizing the computational cost across the set. Moreover, a dual model co-training mechanism extends supervision to unexposed items via mutual pseudo-label refinement, effectively mitigating SSB.
Validated through extensive offline experiments and online A/B testing, DUET consistently outperforms state-of-the-art baselines and achieves improvements across multiple core business metrics. At present, DUET has been fully deployed in Kuaishou and Kuaishou Lite Apps, serving the main traffic for hundreds of millions of users.
\end{abstract}




\maketitle
\begin{figure*}[!t]
    \centering
    \includegraphics[width=1.0\linewidth]{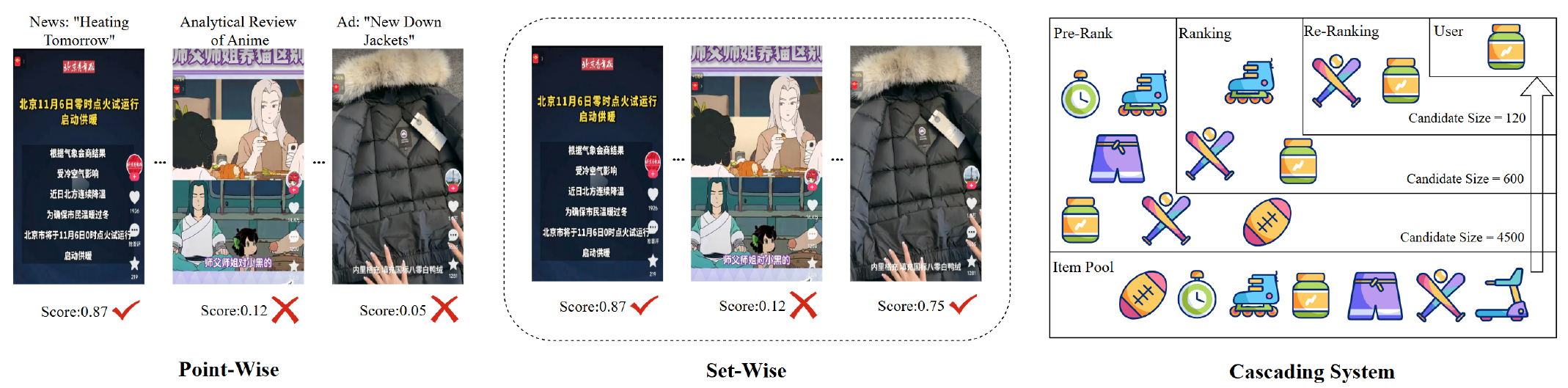}
    \caption{The left and middle panels contrast a point-wise model, which overlooks item synergy, with a set-wise model that captures it. The right panel shows the cascading architecture, where multi-stage filtering creates severe SSB.}
    \label{intro_pic}
\end{figure*}
\section{Introduction}
In industrial recommender systems, the pre-ranking stage must process thousands of candidates under millisecond-level latency. As a result, this stage heavily relies on simple two-tower architectures, which offer low inference cost but suffer from limited estimation capability \cite{chen2025creator,jia2025learn,cao2024moment,su2024rpaf,wang2025findrec}. The computational constraint prevents adopting sophisticated deep models, making pre-ranking the bottleneck of the entire recommendation cascade \cite{qin2022rankflow,wei2024enhancing,zheng2024full}. Consequently, the model’s simplicity not only limits prediction accuracy but also gives rise to two closely coupled challenges that are particularly pronounced at the pre-ranking stage. \cite{chen2024treatment,zhao2025hybrid,chen2024cache}.

Firstly, candidate items often exhibit complex synergistic and suppressive relationships, requiring models to capture inter-item dependencies. In practice, pre-ranking models commonly employ inner-product–based two-tower architectures to cope with time constraints, yet this design sacrifices expressive power and aggravates the limitations of point-wise scoring \cite{zhou2018deepevolutionnetworkclickthrough,guo2017deepfm,si2024twin}. As illustrated in Figure \ref{intro_pic}, a point-wise prediction model would assign an exceptionally high score to a timely and universally relevant news item (e.g., "Timely News: Beijing Heating to Start Tomorrow"). In contrast, a commercial item such as “Ad:New Down Jacket,” which belongs to the long-tail category, would likely receive a relatively modest score and thus be filtered out. Yet, from a set-level perspective, presenting the advertisement immediately after the heating news creates a strong synergy effect: the news stimulates user awareness of warmth needs, and the ad provides a precise solution. This context–product combination not only improves CTR but also enhances user value and commercial revenue.

Secondly, the Sample Selection Bias (SSB) problem is even more severe in pre-ranking \cite{lin2024mitigating,wang2021combating,gao2023rec4ad,liu2022rating,zhang2025mitigating}. This is because pre-ranking operates at the largest candidate scale yet relies on the simplest model form. As illustrated in Figure \ref{intro_pic}, industrial recommender systems employ a cascading architecture that progressively filters a vast item pool through multiple ranking stages, with only a few items ultimately displayed to users. The feedback used for model training is therefore collected exclusively from this final, heavily filtered exposure set, representing only a minute fraction of the initial candidate space. In contrast, the pre-ranking model must evaluate the entire upstream candidate set—often thousands of items per request—where the vast majority are unexposed and lack labels. This creates a much larger distributional gap between training and inference compared with later stages \cite{chen2023bias,huang2022different,chen2021autodebias}. Moreover, the lightweight two-tower models in pre-ranking lack sufficient expressiveness to infer reliable signals for unseen items, thereby exacerbating bias accumulation across cascaded stages.


Critically, these two challenges are not independent. The point-wise paradigm, by isolating item scoring, reinforces users’ inherent preferences and traps them in information cocoons, thereby intensifying SSB. Existing studies remain confined to this paradigm, and both major categories of SSB methods still suffer from fundamental limitations. The first category is grounded in causal inference, such as Inverse Propensity Scoring (IPS) \cite{xu2022dually,ovaisi2020correcting,schnabel2016recommendations} and Doubly Robust (DR) methods \cite{dai2022generalized,li2023multiple,li2022tdr}, aiming to obtain unbiased estimates from exposed samples. However, their scope is strictly limited to the exposed space, overlooking the abundant signals contained in the vast unexposed data. The second category seeks to exploit unexposed samples by generating pseudo-labels through techniques such as knowledge distillation and domain adaptation \cite{xu2022ukd,wei2024enhancing,li2025unbiased}. Nevertheless, these methods typically depend on a single model to generate labels, which not only introduces substantial noise but also lacks effective error-correction mechanisms.

To address the aforementioned challenges, we propose \textbf{DUET} (\textbf{DU}al Model Co-Training for \textbf{E}ntire Space C\textbf{T}R Prediction), a novel and practical framework that collaboratively tackles these two fundamental issues. Our model introduces two key innovations: (i) Our framework departs from conventional two-tower designs and employs a set-wise scoring mechanism that simultaneously captures intra-candidate dependencies and alleviates the computational burden. (ii) We introduce a dual model co-training mechanism that extends the supervisory signal to unexposed samples by interactively generating and correcting pseudo labels across the entire candidate set, thereby mitigating Sample Selection Bias.

Extensive offline experiments and online A/B tests show that DUET consistently surpasses SOTA baselines and improves key business metrics at Kuaishou. It is now fully deployed in both the Kuaishou and Kuaishou Lite apps, serving hundreds of millions of users. To summarize, we highlight the key contributions of this paper as follows:
\begin{itemize}[leftmargin=*, itemsep=0pt, topsep=0pt]
\item \textbf{Framework:} We propose DUET, a practical pre-ranking framework that  adopts a set-wise scoring paradigm to achieve intra-candidate awareness under strict latency budgets, jointly tackles the intertwined challenges of set-level context modeling and SSB. 
\item \textbf{Methodology:} We develop an efficient and stable set-level interaction module to model item dependencies, enabling list-wise optimization under strict latency constraints. And introduce a dual model co-training mechanism to mitigate SSB by extending supervision to unexposed data with robust pseudo-labels.
\item \textbf{Experiment:} Extensive offline and online A/B experiments validate DUET consistently outperforming SOTA methods and improving key business metrics. The framework has been fully deployed in both kuaishou and kuaishou lite app, supporting core traffic for hundreds of millions of users.
\end{itemize}
\section{Preliminary}
In this section, we will briefly introduce the problem definition used in this paper, in order to aid in the reader's understanding.
\subsection{Problem Formulation}\label{subsec:hyperbolic_geometry}
For a given user $u$, we consider a candidate set $C_u = \{ i_1, i_2, \dots, i_m \}$ returned from the retrieval stage, together with a user behavior sequence $S_u = \{ j_1, j_2, \dots, j_n \}$. In conventional point-wise CTR prediction models such as DIN \cite{zhou2018deep}, each candidate item $i \in C_u$ is treated independently. The prediction score, i.e., the click-through probability, is computed based on the user-side features $f_u$ (e.g., user ID, age, gender), item-side features $f_i$ (e.g., item ID, category, author ID), cross features $f_{ui}$, and the user behavior sequence $\{ f_j \}_{j \in S_u}$:
\begin{equation}
    \hat{p}_{u,i} = \mathrm{DIN}(f_u, f_i, f_{ui}, \{ f_j \}_{j \in S_u}), \quad i \in C_u.
\end{equation}

This independent scoring overlooks the interactions among items when presented together, whereas an effective recommendation list should form a coherent ``content ecosystem'' rather than a mere collection of high-scoring items. To this end, we adopt a candidate-set level modeling paradigm: the entire candidate set $\{ (f_i, f_{ui}) \}_{i \in C_u}$, user features $f_u$, and behavior sequence $\{ f_j \}_{j \in S_u}$ are jointly fed into the model, which outputs all click probabilities in a single forward pass:
\begin{equation}
    \hat{\boldsymbol{p}}_{u} = \sigma\!\Big( \mathcal{M}(f_u, \{(f_i,f_{ui})\}_{i \in C_u}, \{ f_j \}_{j \in S_u}; \Theta ) \Big) \in (0,1)^{|C_u|},
\end{equation}
where $\mathcal{M}(\cdot)$ denotes the prediction model parameterized by $\Theta$, $\sigma(\cdot)$ is the element-wise sigmoid function, and $\hat{\boldsymbol{p}}_{u}$ is the predicted probability vector for the candidate set, with its $i$-th component $\hat{p}_{u,i}$ representing the click probability of item $i \in C_u$.
\section{Framework}
\begin{figure*}
\centering
  \includegraphics[scale=.33, trim={0mm 0mm 0mm 0mm}]{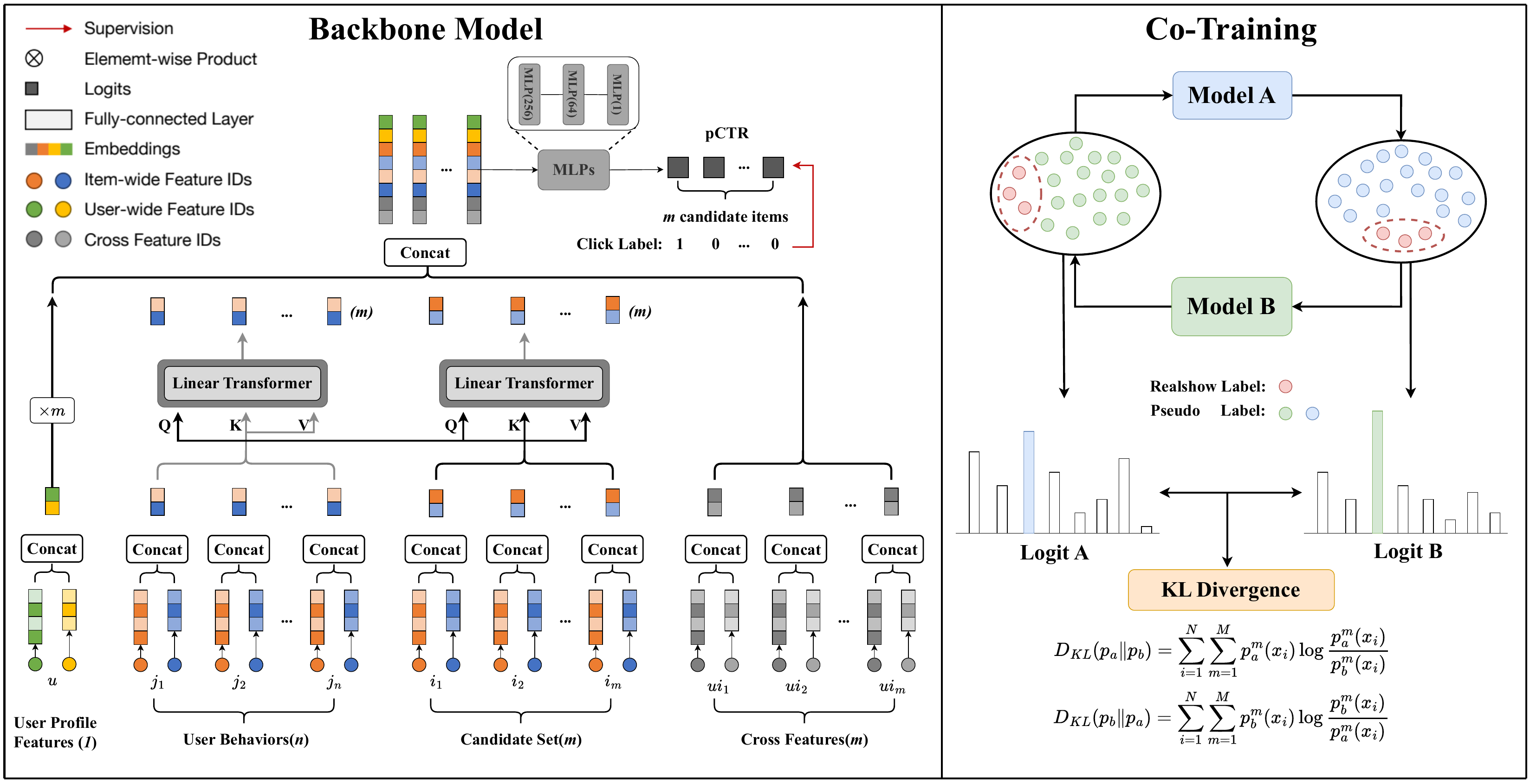}
      \vspace{0mm} 
    \caption{An overview illustration of the DUET architecture.}
    \label{Fig_model}
\end{figure*}

To address the intertwined challenges of inadequate intra-candidate relationship modeling and severe Sample Selection Bias (SSB) that prevail in the pre-ranking stage of large-scale recommender systems, we propose a novel framework, DUET. As depicted in Figure \ref{Fig_model}, the DUET architecture is composed of two core components: a set-level modeling module to capture item interactions, and a dual model co-training mechanism to mitigate SSB. We will elaborate on each component in the following sections.
\subsection{Set-Level Modeling}
In large-scale recommender systems, traditional ranking models typically adopt a point-wise paradigm, where each candidate item is independently modeled to predict its click-through rate (CTR). However, this approach ignores the potential dependencies among items within the candidate set, often resulting in a recommendation list that is a simple stack of high-score items, lacking global consistency and diversity. To address this issue, we propose a set-level modeling approach, which simultaneously captures user--candidate interactions as well as intra-candidate relations to enhance the overall quality of recommendation results.

\noindent\textbf{Feature Representation.}
For a given user $u$, the inputs consist of their profile features $\mathbf{f}_u$, a candidate set $\mathcal{C}_u = \{i_1, \dots, i_m\}$ of $m$ items, and a historical behavior sequence $\mathcal{S}_u = \{j_1, \dots, j_n\}$ of length $n$. We first map all high-dimensional sparse categorical features into low-dimensional dense vector representations via an embedding layer. This process yields four key feature matrices for subsequent interaction modeling:
\begin{itemize}[leftmargin=*]
    \item \textbf{Candidate Embedding Matrix} $\mathbf{F}_{\text{can}} \in \mathbb{R}^{m \times d}$: Stacked embeddings of the $m$ candidate items.
    \item \textbf{Sequence Embedding Matrix} $\mathbf{F}_{\text{seq}} \in \mathbb{R}^{n \times d}$: Stacked embeddings of the $n$ historical behavior items.
    \item \textbf{User Profile Embedding Matrix} $\mathbf{F}_{u} \in \mathbb{R}^{m \times d_u}$: The user profile vector $\mathbf{f}_u$ tiled $m$ times to match the candidate set dimension.
    \item \textbf{Cross Feature Embedding Matrix} $\mathbf{F}_{\text{cro}} \in \mathbb{R}^{m \times d_c}$: Embeddings of user-item cross features for each candidate.
\end{itemize}

\noindent\textbf{User-Candidate Interaction Modeling.}
To accurately capture the alignment between a user's dynamic interests and the candidate items, we must model the deep interactions between the candidate set $\mathbf{F}_{\text{can}}$ and the user's long behavior sequence $\mathbf{F}_{\text{seq}}$. A standard attention mechanism, with $\mathbf{F}_{\text{can}}$ as the Query and $\mathbf{F}_{\text{seq}}$ as the Key and Value, would incur a computational complexity of $O(mn)$. This is computationally prohibitive in industrial scenarios where both $m$ and $n$ can be in the thousands or larger.

To achieve efficient full-sequence interaction, we incorporate the Linear Attention mechanism \cite{katharopoulos2020transformers}. By replacing the softmax function with a kernel function $\phi(\cdot)$ and leveraging the associative property of matrix multiplication, the computation order can be rearranged from $(\phi(\mathbf{F}_{\text{can}}) \phi(\mathbf{F}_{\text{seq}})^{\top}) \mathbf{F}_{\text{seq}}$ to $\phi(\mathbf{F}_{\text{can}}) (\phi(\mathbf{F}_{\text{seq}})^{\top} \mathbf{F}_{\text{seq}})$. This transformation reduces the complexity from quadratic, $O(mn)$, to linear, $O(m+n)$. The representation $\mathbf{F}_{\text{c-s}}$ is defined as:
\begin{equation}
    \mathbf{F}_{\text{c-s}} = \text{LinearAttn}(\mathbf{F}_{\text{can}}, \mathbf{F}_{\text{seq}}, \mathbf{F}_{\text{seq}}) \in \mathbb{R}^{m \times d},
    \label{eq:cross_attn}
\end{equation}
where each row in $\mathbf{F}_{\text{c-s}}$ aggregates the interest representation for the corresponding candidate item, distilled from the user's history.

\noindent\textbf{Intra-Candidate Relation Modeling.}
Beyond user-item matching, a high-quality recommendation list must account for the intrinsic relationships among items. For instance, a news item followed by a contextually relevant advertisement can create a synergistic effect, whereas presenting multiple homogeneous items may lead to user fatigue. To capture these contextual dependencies, we design a second attention stream that performs self-attention within the candidate set. We again employ Linear Attention to ensure computational efficiency. This process is formalized as:
\begin{equation}
    \mathbf{F}_{\text{c-c}} = \text{LinearAttn}(\mathbf{F}_{\text{can}}, \mathbf{F}_{\text{can}}, \mathbf{F}_{\text{can}}) \in \mathbb{R}^{m \times d}.
    \label{eq:self_attn}
\end{equation}

The matrix $\mathbf{F}_{\text{c-c}}$ encodes the contextual information of each item within the current candidate "ecosystem," enabling the model to perceive collaborative and competitive relationships.

\noindent\textbf{Fusion and Prediction.}
Finally, we concatenate the outputs from the two attention modules, $\mathbf{F}_{\text{c-s}}$ and $\mathbf{F}_{\text{c-c}}$, with the user profile embeddings $\mathbf{F}_{u}$ and cross feature embeddings $\mathbf{F}_{\text{cro}}$. This forms a final comprehensive feature representation matrix $\mathbf{F}_{\text{final}}$:
\begin{equation}
    \mathbf{F}_{\text{final}} = \text{Concat}(\mathbf{F}_{\text{c-s}}, \mathbf{F}_{\text{c-c}}, \mathbf{F}_{u}, \mathbf{F}_{\text{cro}}) \in \mathbb{R}^{m \times d_{\text{final}}}.
    \label{eq:concat}
\end{equation}

This matrix is then fed into a Multi-Layer Perceptron (MLP) for non-linear transformation. A final Sigmoid activation function is applied to produce the predicted click-through rates (pCTR) for all $m$ candidate items simultaneously.
\begin{equation}
    \hat{\mathbf{y}} = \sigma(\text{MLP}(\mathbf{F}_{\text{final}})) \in \mathbb{R}^{m \times 1}
    \label{eq:prediction}
\end{equation}
Through this set-wise modeling approach, our model not only predicts the individual relevance of each item but also optimizes the global utility of the entire recommendation list.

\noindent\textbf{Degree Normalization.} While Linear Attention offers significant efficiency gains, the removal of the softmax function also eliminates its inherent normalization property. This can lead to numerical explosion during the aggregation process, causing training instability.

To address this issue, we draw inspiration from the normalization strategy employed in Graph Convolutional Networks (GCNs) \cite{yu2021self}. GCNs normalize aggregated features by the node degree to prevent features from high-degree nodes from dominating. Analogously, we can conceptualize the sum of similarities between a query and all keys as its "attention degree."

We introduce a degree normalization term to stabilize the training process. Specifically, the Linear Attention computation in Equation~\ref{eq:cross_attn} is reformulated as:
\begin{equation}
    \mathbf{E} = \mathbf{D}^{-1} \phi(\mathbf{Q}) (\phi(\mathbf{K})^{\top}\mathbf{V}),
    \label{eq:gcn_norm}
\end{equation}
where $\mathbf{Q}=\mathbf{F}_{\text{can}}$, and $\mathbf{K}=\mathbf{V}=\mathbf{F}_{\text{seq}}$. Here, $\mathbf{D}$ is a diagonal matrix whose $i$-th diagonal element is the sum of similarities between the $i$-th query and all keys, defined as $\mathbf{D} = \text{diag}(\phi(\mathbf{Q})\phi(\mathbf{K})^{\top}\mathbf{1})$, with $\mathbf{1}$ being a column vector of ones. This degree normalization effectively regularizes the aggregation, ensuring training stability while preserving the linear complexity benefits.

\subsection{Co-Training}
To effectively mitigate the severe Sample Selection Bias (SSB) problem arising from training solely on exposed feedback, we introduce a dual model co-training mechanism. The core of this mechanism lies in utilizing two independent models to extend the supervisory signal from the sparse exposed space to the entire unexposed candidate space through interactive pseudo-labeling.

Specifically, the framework comprises two models with identical architectures but independently initialized parameters, denoted as $M_A(\cdot; \theta_A)$ and $M_B(\cdot; \theta_B)$. For any given batch of candidates $\mathcal{D}$, we partition it into two subsets: the exposed set $\mathcal{D}_{obs}$, which contains ground-truth labels $\{y_i\}$, and the unexposed set $\mathcal{D}_{unobs}$, which lacks labels. The co-training mechanism operates as follows:

\noindent\textbf{Interactive Pseudo-Labeling.} For samples $x_i$ in the unexposed set $\mathcal{D}_{unobs}$, the two models act as mutual teachers to provide supervisory signals. Specifically, the learning target for model $M_A$ (i.e., the pseudo-label) is given by the prediction of model $M_B$, and vice versa. This process is formalized as:
\begin{equation}
    \tilde{y}_{i}^A = M_B(x_i; \theta_B) \quad \text{and} \quad \tilde{y}_{i}^B = M_A(x_i; \theta_A), \quad \forall x_i \in \mathcal{D}_{unobs}
\end{equation}
\noindent where $\tilde{y}_{i}^A$ and $\tilde{y}_{i}^B$ are the soft pseudo-labels used to train $M_A$ and $M_B$, respectively. Since the models learn from different initial states, they capture distinct aspects and biases of the data, forming complementary "views". This view diversity enables one model to correct potential errors made by the other in its pseudo-label generation, establishing a dynamic and adaptive error-correction loop that effectively suppresses the confirmation bias common in single-model pseudo-labeling approaches.

\noindent\textbf{Consistency Regularization.}
While view diversity is crucial for the error-correction mechanism, excessive divergence can lead to noisy pseudo-labels and destabilize the training process. To strike a balance between diversity and consensus, we introduce a consistency regularization term that penalizes discrepancies between the predictive distributions of the two models. We treat the model outputs as parameters of a Bernoulli distribution and employ the symmetric Kullback-Leibler (KL) divergence to measure the difference between these two distributions, $P_A$ and $P_B$:
\begin{equation}
    \mathcal{L}_{\text{con}} = \mathbb{E}_{x \in \mathcal{D}} \left[ \text{KL}(P_A(\cdot|x) || P_B(\cdot|x)) + \text{KL}(P_B(\cdot|x) || P_A(\cdot|x)) \right].
\end{equation}
\subsection{Optimization}
For model $M_A(\cdot; \theta_A)$, its objective is composed of two main components. The first is the supervised loss, $\mathcal{L}_{\text{sup}}^A$, which employs the BCE loss function. The source of supervision for this loss depends on the sample's origin. For samples in the exposed set $\mathcal{D}_{obs}$, we use the ground-truth label $y_i$. Conversely, for samples in the unexposed set $\mathcal{D}_{unobs}$, we use the soft pseudo-label $\hat{y}_i^B = \sigma(M_B(x_i; \theta_B))$ generated by model $M_B$, where $\sigma(\cdot)$ is the Sigmoid activation function. We define this conditional loss as follows:
\begin{equation}
    \mathcal{L}_{\text{sup}}^A = - \mathbb{E}_{x_i \in \mathcal{D}} \left[ \begin{cases} \mathcal{L}_{\text{BCE}}(y_i, \hat{y}_i^A) & \text{if } x_i \in \mathcal{D}_{obs} \\ \mathcal{L}_{\text{BCE}}(\hat{y}_i^B, \hat{y}_i^A) & \text{if } x_i \in \mathcal{D}_{unobs} \end{cases} \right].
\end{equation}

The second part is the consistency regularization term, $\mathcal{L}_{\text{consist}}$, as defined previously, which aims to align the predictive distributions of the two models. Combining these two parts, the final optimization objective for model $M_A$, denoted as $\mathcal{L}_A$, is defined as:
\begin{equation}
    \mathcal{L}_A(\theta_A, \theta_B) = \mathcal{L}_{\text{sup}}^A + \lambda \mathcal{L}_{\text{con}},
\end{equation}
where $\lambda$ is a hyperparameter that balances the importance of fitting the supervisory signals against maintaining inter-model consistency. The optimization objective for model $M_B$, $\mathcal{L}_B(\theta_A, \theta_B)$, is defined in a perfectly symmetric manner.
\subsection{Deployment of DUET}
\begin{figure}[ht]
    \centering
\includegraphics[width=1\linewidth]{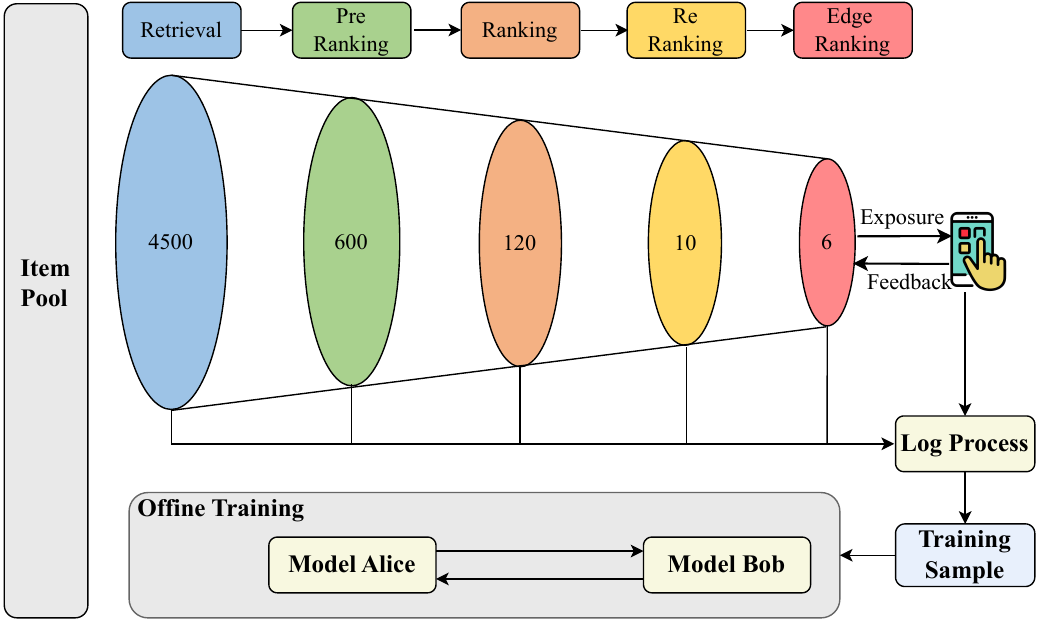}
    \caption{The deployment of DUET at Kuaishou.}
    \label{deployment}
\end{figure}

DUET is deployed in Kuaishou's production recommender system, serving hundreds of millions of users. As shown in Figure \ref{deployment}, the deployment consists of two main components: the Online Serving Architecture and the Offline Training Pipeline.
\paragraph{Online Serving Architecture}
Industrial recommender systems typically employ a multi-stage cascaded architecture to reconcile the trade-off between recommendation quality and system latency. Our framework DUET is deployed at the pre-ranking stage. Positioned after retrieval, this stage is tasked with efficiently refining an initial candidate pool of 4,500 items into a higher-value subset of 600. This curated subset is then passed to subsequent modules for the final ranking presented to the user.
\paragraph{Offline Training Pipeline}
The offline training pipeline of DUET operates as an automated cyclical process with the following stages:
\begin{enumerate}[leftmargin=*]
    \item \textbf{Log Processing:} The process commences with a log processing module that collects real-time user interactions (e.g., clicks, watch time) with exposed items. These interactions form the basis for generating labeled training samples.
    \item \textbf{Entire Space Co-Training:} To mitigate the severe Sample Selection Bias (SSB) from training only on exposed items, we incorporate unexposed data. The framework's dual models, Alice and Bob, perform corrective co-training by interactively generating and refining pseudo-labels for each other on these samples. This mutual supervision enhances model generalization across the entire candidate space.
    \item \textbf{Model Update and Deployment:} Upon completion of the training cycle, the superior performing model is deployed to the online serving environment, replacing the previous pre-ranking model. Notably, only one model is deployed to preserve efficiency and avoid redundant serving cost.
\end{enumerate}
\vspace{-5pt}
\section{Experiments}
To validate the performance of our DUET framework, we design and conduct comprehensive experimental studies. In this section, we present our experimental setup and results analysis. Specifically, our experiments aim to investigate the following research questions:
\begin{itemize}[leftmargin=*]
\item\textbf{RQ1}: How does DUET perform in offline experiments on real-world datasets compared to baseline models?
\item\textbf{RQ2}: How does each proposed module contribute to the performance?
\item\textbf{RQ3}: How do hyper-parameters influence the performance of
DUET?
\item\textbf{RQ4}: How does DUET perform in real online CTR Prediction?
\end{itemize}
\subsection{Experimental Setup}
\subsubsection{Datasets.}
We validate our approach on the Recflow dataset and further evaluate it using streaming data from Kuaishou’s production environment, demonstrating robustness in both academic and industrial contexts. Detailed statistics are provided in Table \ref{dataset_sta}.

\textbf{Public Dataset:} Since this study requires datasets that include candidate sets across the entire recommendation pipeline, to the best of our knowledge, RecFlow \cite{liu2024recflow} is the only publicly available dataset that meets this requirement. RecFlow provides comprehensive coverage of exposed and unexposed samples across all stages of the recommendation pipeline. Following prior work [44, 45], we adopt the leave-one-out strategy to partition each user's interactions into training, validation, and test sets.

\textbf{Private Dataset:} To further validate the effectiveness of our model in real-world industrial settings, we train and evaluate the model directly on streaming data from Kuaishou’s online recommendation system to assess its performance in large-scale industrial environments. The dataset is collected from the real-time data streams of Kuaishou’s online recommendation system. It continuously receives exposure and click feedback from real users and covers the entire recommendation pipeline, ranging from candidate retrieval to final exposure. 
\vspace{-5pt}
\begin{table}[h]
\centering
\caption{The statistics of datasets (M: million, B: billion).}
\setlength{\tabcolsep}{0.55mm}{ 
\begin{tabular}{lcccccc}
\toprule
\textbf{Dataset} & \textbf{User} & \textbf{Item} & \textbf{Sample} & \textbf{Request} & \textbf{Impression} & \textbf{Click} \\
\midrule
Recflow & 42471 & 5M & 524M & 1.1M & 4.8M & 2.1M \\
Industrial & 144M & 71M & 75B & 300M & 980M & 608M \\
\bottomrule
\end{tabular}
}
\label{dataset_sta}
\end{table}
\vspace{-5pt}
\subsubsection{Evaluation Metrics.}
To evaluate model performance, we use
AUC (Area Under the Curve) and UAUC (User-level AUC) as the primary performance metrics.
Following~\cite{zhou2018deep}, we adopt the \textit{Relative Improvement} (RelaImpr) metric to evaluate model performance improvements. Considering a random guesser yields an AUC of 0.5, RelaImpr is defined as:
\begin{equation}
\text{RelaImpr} = \left( \frac{\text{AUC}(\text{measured model}) - 0.5}{\text{AUC}(\text{base model}) - 0.5} - 1 \right) \times 100\%.
\end{equation}
\subsubsection{Baselines.}
To ensure a thorough assessment, the baselines are organized into two distinct categories.

\noindent\textbf{First, Causal Methods.} The first group of methods aims to mitigate sample selection bias by integrating causal inference.
\begin{itemize}[leftmargin=*]
    \item \textbf{Multi-DR} \cite{zhang2020large}: This method leverages multi-task learning to combine causal modeling and error imputation, thereby ensuring robustness to propensity misspecification.
    \item \textbf{Multi-IPS} \cite{zhang2020large}: This method employs inverse propensity weighting within a multi-task learning framework to produce unbiased conversion rate estimates given accurate propensities.
    \item \textbf{DR-JL} \cite{wang2019doubly}: This method introduces a principled doubly robust approach to alleviate bias from inaccurate imputations and variance from propensity estimation.
    \item \textbf{MRDR} \cite{guo2021enhanced}: This method integrates theoretical variance reduction with empirical joint learning to achieve stable and reliable debiasing under missing not at random conditions.
    \item \textbf{DCMT} \cite{zhu2023dcmt}: This method formulates conversion prediction as a causal multi-task problem and directly incorporates counterfactual samples to correct hidden negative biases.
    \item \textbf{DDPO} \cite{su2024ddpo}: This method combines causality-based optimization with dynamic soft-labeling, thereby enhancing generalization and robustness across both observed and unobserved samples.
\end{itemize}
\noindent\textbf{Second, Empirical Methods.} This group of methods mitigates exposure bias by generating pseudo-labels for unexposed samples.
\begin{itemize}[leftmargin=*]
    \item \textbf{UKD} \cite{xu2022ukd}: This method formulates conversion prediction as a knowledge distillation task, explicitly incorporating unclicked samples with uncertainty modeling to enhance robustness.
    \item \textbf{SIDA} \cite{wei2024enhancing}: This method adopts a domain adaptation module that generates exposure-independent pseudo labels to enhance robustness against exposure bias.
    \item \textbf{UECF} \cite{li2025unbiased}: This method generates reliable pseudo-labels for unexposed samples through feature-level disentanglement, thereby mitigating exposure bias.
\end{itemize}
\subsubsection{Implementation Details.}
We select the optimal hyperparameters based on the AUC metric evaluated on the validation set. The maximum sequence length is fixed at 200, and the embedding dimension is uniformly set to 64 across both datasets. The AdamW optimizer is employed, with the learning rate chosen from $\{1\mathrm{e}{-3},\, 5\mathrm{e}{-3},\, 1\mathrm{e}{-4},\, 5\mathrm{e}{-4},\, 1\mathrm{e}{-5},\, 5\mathrm{e}{-5}\}$
. All other hyperparameters are kept consistent with the original configurations reported in their respective studies. We adopt the widely used DIN model as the base CTR prediction framework \cite{zhou2018deepevolutionnetworkclickthrough,xiao2025marsmodalityalignedretrievalsequence}. All experiments are implemented in TensorFlow and executed on a computing environment equipped with 25 Intel Xeon CPU cores (2.10 GHz) and two NVIDIA RTX A10 GPUs, providing sufficient computational resources for large-scale training and hyperparameter optimization.
\subsection{Overall Performance (RQ1)}
Table \ref{main_table} reports the comprehensive performance of all the compared baselines across Recflow and Industrial datasets. Based on the results, the main
observations are as follows:
\begin{itemize}[leftmargin=*]
\item Compared with baseline methods in the category of causal inference approaches (e.g., Multi-IPS, DR-JL, MRDR), DUET demonstrates significant and consistent improvements across all evaluation metrics. While these causal estimators aim to achieve unbiased learning from observed exposure data, their core design is restricted to bias correction within the observed sample space, thereby failing to exploit the rich signals inherent in unexposed data. Although recent methods such as DCMT and DDPO attempt to incorporate unobserved samples, they still derive propensity scores solely from observed data and neglect the necessity of separately modeling the unobserved space. As a result, residual bias is introduced, ultimately limiting their effectiveness.
\item Within the family of empirical baselines, DUET consistently demonstrates superior performance. While these empirical methods attempt to leverage unexposed samples, they typically rely on a single model to generate pseudo labels. This paradigm suffers from inherent limitations: it is prone to introducing substantial noise, resulting in pseudo labels of uneven quality, and lacks effective error-correction mechanisms to ensure the generation of high quality pseudo-labels. In contrast, the Co-Training framework of DUET establishes a dynamic error-correction loop through mutual supervision and information exchange between two models, thereby producing pseudo labels that are both higher in quality and more robust.
\item DUET's consistently superior performance across diverse datasets demonstrates its robust generalization and industrial applicability. This robustness stems from systematically addressing two intertwined challenges: SSB and the absence of intra-candidate interaction modeling. DUET's dual model co-training mechanism enables mutual supervision between models, generating high-quality pseudo-labels for unexposed samples. Additionally, its set-level modeling explicitly captures synergistic or suppressive relationships among candidates, which are overlooked by traditional point-wise methods. By capturing these interactions, DUET enhances recommendation lists' global coherence and diversity, proving particularly critical in industrial scenarios with large-scale, complex candidate sets.
\end{itemize}
\begin{table}[h]
\centering
\caption{The overall performance evaluation results of the proposed method and compared methods on two benchmark datasets. Results are grouped into heuristic and trainable methods. Within each group, the best and second-best performances are highlighted in bold and borderline, respectively. Numbers with an asterisk (*) indicate statistically significant improvements over the best trainable baseline (t-test with p-value < 0.05).}
\setlength{\tabcolsep}{1.2pt} 
\begin{tabular}{l|cc|cccc}
\toprule
\multirow{2}{*}{Model} &
\multicolumn{2}{c|}{Recflow} &
\multicolumn{4}{c}{Industrial} \\
\cmidrule(lr){2-3}\cmidrule(lr){4-7}
& AUC & RelaImpr & AUC & RelaImpr & UAUC & RelaImpr \\
\midrule
BaseModel & 0.6127 & NA & 0.6730 & NA & 0.6675 & NA \\
\midrule
\multicolumn{7}{c}{\textbf{Causal Estimation
Scenario}}\\
\midrule
Multi-IPS      & 0.6141 & 1.24\% & 0.6755 & 1.47\% & 0.6679 & 0.27\% \\
Multi-DR   & 0.6133 & 0.53\% & 0.6751 & 1.24\% & 0.6677 & 0.12\% \\
DR-JL    & \underline{0.6189} & \underline{5.50\%} & 0.6769 &  2.28\% & 0.6687 & 0.74\% \\
MRDR      & 0.6160 & 2.92\% & 0.6788 & 3.36\% & 0.6693 & 1.09\% \\
DCMT    & 0.6173 & 4.08\% & 0.6801 & 4.15\% & 0.6715 & 2.43\% \\
DDPO    & 0.6179 & 4.61\% & \underline{0.6826} & \underline{5.56\%} & \underline{0.6718} & \underline{2.61\%} \\
\textbf{OURS} & \textbf{0.6322$^{\ast}$} & \textbf{17.31\%} & \textbf{0.6881$^{\ast}$} & \textbf{8.73\%} & \textbf{0.6787$^{\ast}$} & \textbf{6.69\%} \\
\midrule
\multicolumn{7}{c}{\textbf{Empirical Estimation Scenario}}\\
\midrule
UKD  & 0.6171 & 3.91\% & 0.6812 & 4.76\% & 0.6694 & 1.19\% \\
SIDA  & 0.6235 & 9.58\% & 0.6824 & 5.44\% & 0.6731 & 3.36\% \\
UECF   & \underline{0.6273} & \underline{12.96\%} & \underline{0.6836} & \underline{6.12\%} & \underline{0.6756} & \underline{4.83\%} \\
\midrule
\textbf{OURS} & \textbf{0.6322$^{\ast}$} & \textbf{17.31\%} & \textbf{0.6881$^{\ast}$} & \textbf{8.73\%} & \textbf{0.6787$^{\ast}$} & \textbf{6.69\%} \\
\bottomrule
\end{tabular}
\label{main_table}
\end{table}
\subsection{Ablation Study (RQ2)}
To evaluate the effectiveness of the proposed DUET framework, we conducted a series of ablation studies by removing three key components: the set-level interaction module (DUET w/o Set), the dual model co-training mechanism (DUET w/o Co), and the distribution alignment regularizer (DUET w/o KL). Table \ref{ablation} presents our results, from which we derive the following important conclusions.
\begin{itemize}[leftmargin=*]
\item Among all ablations, removing the dual model co-training mechanism (w/o Co) leads to the most pronounced performance degradation. This highlights the pivotal role of co-training in addressing the SSB problem. Without this mechanism, the framework degenerates into a single model pseudo labeling paradigm, which is highly susceptible to confirmation bias and thus fails to generate reliable and high quality labels. By contrast, DUET leverages a dual model architecture in which two independently parameterized models mutually supervise each other, forming a dynamic error correction loop. The design guarantees the reliability and robustness of the generated pseudo labels, thereby strengthening the model’s generalization across the entire data space.
\item Removing the set-level interaction module (w/o Set) also causes a notable decline in performance. This finding underscores the necessity of explicitly modeling the complex dependencies among items within a candidate set during the pre-ranking phase. Traditional point-wise methods evaluate candidates in isolation, overlooking the synergistic or suppressive effects that arise when items are jointly presented. In contrast, DUET’s set-level module leverages a self-attention mechanism to capture such contextual information, transforming the prediction task from independent item scoring into a holistic list-wise optimization problem. This enables the model to generate more coherent and diverse recommendations that better align with user preferences.
\item Ablating the distribution alignment regularization term (w/o KL) impairs model performance. The essence of co-training lies in leveraging the complementary perspectives of two independent models for mutual correction and reinforcement. However, this view discrepancy is a double-edged sword. On one hand, it generates diverse supervisory signals, preventing either model from converging to a local optimum. On the other hand, if left unconstrained, it can exacerbate predictive divergence between the models, leading to conflicting pseudo-labels and ultimately disrupting the collaborative mechanism. To balance this trade-off, We introduce KL divergence as a soft constraint not to eliminate these model discrepancies, but to guide them. 
\end{itemize}
\vspace{-5pt}
\begin{table}[h]
\centering
\caption{Performance of design variants on two datasets. Bold numbers denote the largest performance changes.}
\label{xiaorong}
\setlength{\tabcolsep}{2pt}
\begin{tabular}{c|cc|cc|cc}
\toprule
\multirow{2}{*}{Model} & \multicolumn{2}{c|}{Recflow} & \multicolumn{4}{c}{Industrial}  \\
\cmidrule(lr){2-3} \cmidrule(lr){4-7}
& AUC & RelaImpr & AUC & RelaImpr & UAUC & RelaImpr \\
\midrule
DUET & 0.6322 & 0.00\% & 0.6881 & 0.00\% & 0.6787 & 0.00\% \\
w/o Set & 0.6279 & -3.36\% & 0.6813 & -3.75\% & 0.6732 & -3.16\% \\
w/o Co & \textbf{0.6257} & \textbf{-5.17\%} & \textbf{0.6792} & \textbf{-4.97\%} & \textbf{0.6703} & \textbf{-4.93\%} \\
w/o KL & 0.6306 & -1.22\% & 0.6834 & -2.56\% & 0.6749 & -2.17\% \\
\bottomrule
\end{tabular}
\label{ablation}
\end{table}
\vspace{-5pt}
\subsection{In-depth Analysis (RQ3)}
\begin{figure}[h!]
    \centering
\includegraphics[width=0.47\textwidth]{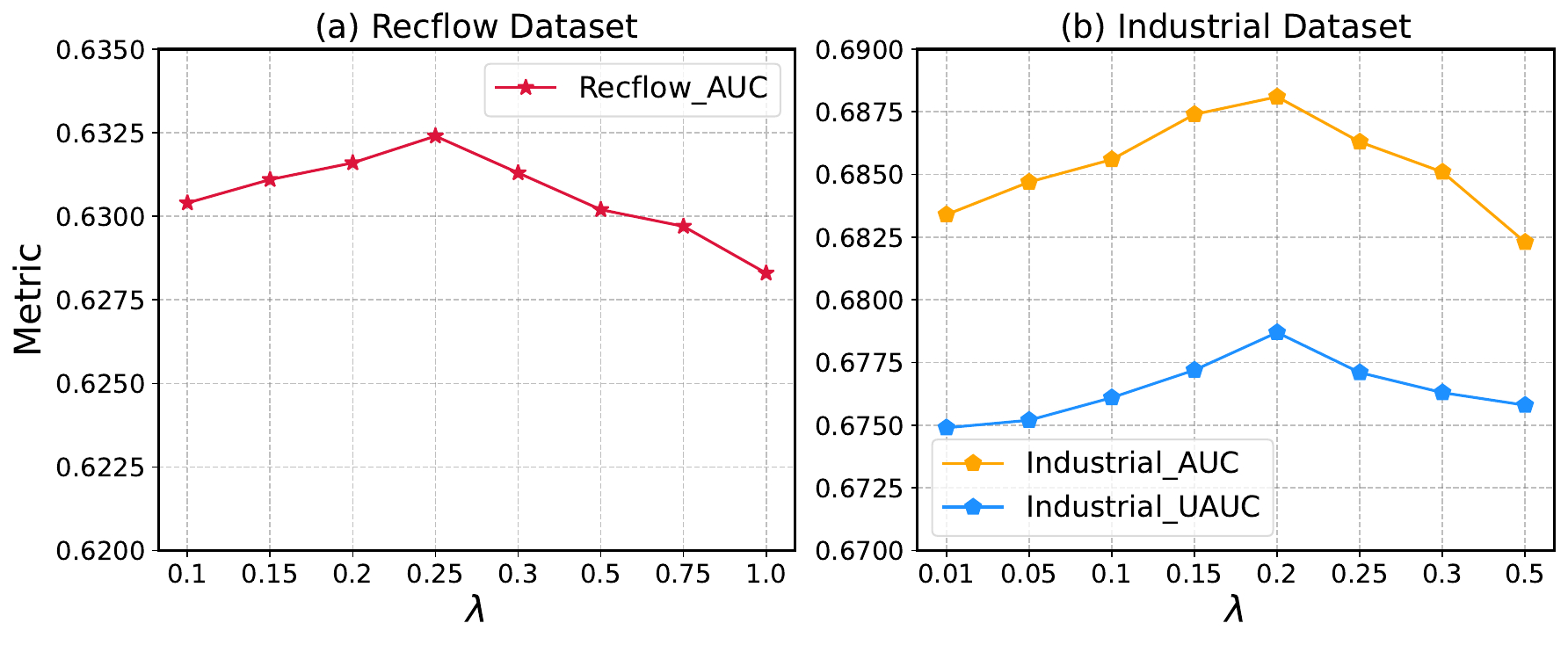}
    \caption{The impact of the regularization strength \(\lambda\).}
    \label{chaocan}
\end{figure}

The regularization coefficient \(\lambda\) governs the essential trade-off between view diversity and prediction consensus in the dual model. We conduct a sensitivity analysis to examine its impact, and the results are shown in Figure~\ref{chaocan}. 

We observe a clear unimodal relationship between \(\lambda\) and model performance, validating the underlying rationale of our design. When \(\lambda\) is too small, the weak constraint allows excessive divergence between the two models. Their predictions drift apart, producing inconsistent pseudo-labels that undermine the co-training mechanism. Conversely, an overly large \(\lambda\) forces the models into homogenization. Although this ensures consistency, it suppresses the diversity needed for mutual correction and causes the paradigm to degenerate into self-training. for mutual correction and causes.
\vspace{-15pt}
\subsection{Online Result (RQ4)}
To evaluate the performance of the DUET framework in a real-world industrial environment, we deployed a two-week online A/B test on both the Kuaishou and Kuaishou Lite applications. As detailed in
Table \ref{online}, the experimental results clearly demonstrate that our proposed model surpasses the baseline across multiple core business metrics, achieving statistically significant improvements.

Specifically, DUET delivered significant growth in user engagement and retention. On the Kuaishou app, for instance, Total Watch Time and Avg. App Usage Time increased by +0.195\% and +0.147\%, respectively. This indicates that by generating more accurate predictions over the entire candidate space and modeling item synergies, DUET produces more compelling recommendation lists, thereby effectively enhancing user retention and consumption depth.

Furthermore, DUET fosters a healthier content ecosystem and drives long-term value. This is evidenced by significant uplifts in diversity-oriented metrics, including Avg. Novel Cluster (+0.173\%) and Avg. Surprise Cluster Exposure (+0.165\%). These results underscore DUET's capacity to surface novel content, thereby mitigating the risk of filter bubbles and broadening user interest boundaries. 
These improvements in content ecology, together with the retention gains, demonstrate that the model delivers tangible and multifaceted value in a large-scale production environment.
\begin{table}[h]
\centering
\caption{A/B test performance on Kuaishou and Kuaishou~Lite across engagement metrics. The Improvement column reports the relative percentage uplift of the treatment over the control. The Confidence Interval (CI) provides a range of plausible values for the true uplift. A CI that does not contain 0 indicates that the result is statistically significant.}
\label{online}
\small 
\begin{tabular}{@{}c lcc@{}} 
\toprule
Scenario & Metric & Improv. & CI \\
\midrule
\multirow{7}{*}{Kuaishou} 
 & Avg. App Usage Time & +0.147\% & [0.07\%, 0.22\%] \\
 & Total Watch Time & +0.195\% & [0.09\%, 0.30\%] \\
 & Video Watch Time & +0.125\% & [0.00\%, 0.25\%] \\
 & Avg. Suprise Cluster & +0.165\% & [0.09\%, 0.24\%] \\
 & Avg. Novel Cluster & +0.173\% & [0.11\%, 0.24\%] \\
 & Avg. Long-Term Cluster Exp. & +0.100\% & [0.03\%, 0.17\%] \\
 & Avg. Accurately Interest & +0.149\% & [0.05\%, 0.24\%] \\
\midrule
\multirow{7}{*}{\begin{tabular}[c]{@{}c@{}}Kuaishou\\ Lite\end{tabular}} 
 & Avg. App Usage Time & +0.049\% & [0.01\%, 0.09\%] \\
 & Total Watch Time & +0.061\% & [0.01\%, 0.11\%] \\
 & Video Watch Time & +0.056\% & [0.01\%, 0.10\%] \\
 & Number of Novel Interests & +0.058\% & [0.01\%, 0.11\%] \\
 & Avg. Long-Term Interest Exp. & +0.041\% & [0.01\%, 0.08\%] \\
 & Avg. Cluster Exp. & +0.051\% & [0.02\%, 0.08\%] \\
 & Avg. Effectively Clusters & +0.050\% & [0.01\%, 0.08\%] \\
\bottomrule
\end{tabular}
\end{table}
\section{Related Work}
In this section, we review two relevant prior works: Click-through
rate model and sample selection bias.

\subsection{Click-Through-Rate Prediction}
Research on CTR prediction progressed from linear and shallow factorization models, including Logistic Regression (LR) \cite{mcmahan2013ad}, Factorization Machines (FM) \cite{rendle2010factorization}, and Field-aware Factorization Machines (FFM) \cite{juan2016field}, to deep architectures that combine memorization and generalization. Wide \& Deep \cite{cheng2016wide}, DeepFM \cite{guo2017deepfm}, and DCN \cite{wang2017deep,wang2021dcn} enable end-to-end learning of higher-order feature interactions. To model evolving interests, the DIN family adopts behavior-sequence modeling and advances from relevance attention to interest evolution, short-term preference modeling, and unified treatment of short and long terms \cite{zhou2018deep,zhou2019deep,feng2019deep,chen2019behavior}. Leveraging extended user histories is also crucial: HPMN \cite{Ren_2019} and MIMN \cite{Pi_2019} use external memory networks to preserve persistent preferences. To reduce the inconsistency between short and long terms, retrieval then modeling frameworks first retrieve salient subsequences and then apply specialized modeling. SIM \cite{qi2020searchbasedusermodelinglifelong} and UBR4CTR \cite{Qin_2020,qin2023learning} follow this design; ETA \cite{chen2021endtoenduserbehaviorretrieval} and SDIM \cite{cao2022samplingneedmodelinglongterm} improve retrieval precision with locality-sensitive hashing and signature mechanisms. TWIN-V2 \cite{si2024twin} scales historical modeling to $10^6$ behaviors using hierarchical clustering with cluster-aware attention, enabling long range retrieval with real time efficiency.
\subsection{Sample Selection Bias}
Sample Selection Bias (SSB) is a critical challenge in recommender systems and online advertising, arising because models are trained on biased data from observed user feedback but must perform inference on the entire, unbiased data space \cite{ma2023selection,chen2023bias,chen2021bias,li2025survey,pan2021correcting}. Causal inference is a primary approach to mitigate Sample Selection Bias (SSB). Initial work employed methods like Inverse Propensity Scoring (IPS) \cite{xu2022dually,allan2020propensity,chen2022best,ma2023selection} and Doubly Robust (DR) estimators \cite{song2023cdr,li2022stabledr,saito2020doubly,zou2022approximated}, as demonstrated in the doubly robust joint learning approach for recommendation. These have been enhanced in frameworks like MRDR \cite{guo2021enhanced} to reduce variance, and in Multi-IPW/Multi-DR \cite{zhang2020large} to combat data sparsity via multi-task learning. Advanced methods extend this by leveraging the entire data space, often by generating pseudo-labels for unobserved samples. This is achieved through techniques such as the dual propensity networks in DDPO \cite{su2024ddpo}, the counterfactual framework in DCMT \cite{zhu2023dcmt}, and the uncertainty-regularized knowledge distillation in UKD \cite{xu2022ukd}. Specifically for the challenging pre-ranking stage, comprehensive frameworks like SIDA \cite{wei2024enhancing} combine sample selection with unbiased distillation, while UECF \cite{li2025unbiased} introduces an unbiased causal framework with automatic sample filtering to ensure theoretical unbiasedness.
\section{Conclusion}
This paper addresses the critical, intertwined challenges of Sample Selection Bias (SSB) and point-wise scoring limitations in the pre-ranking stage of recommender systems. We proposed DUET, a novel framework that integrates a dual model co-training mechanism with an efficient set-level interaction module. The co-training paradigm effectively mitigates SSB by extending supervision to the entire candidate space via interactive pseudo-label generation and error correction. Simultaneously, the set-level module captures complex inter-item dependencies, enabling holistic list-wise optimization. Extensive offline experiments and large-scale online A/B tests have validated DUET's significant superiority over state-of-the-art baselines. Its successful deployment in Kuaishou, serving hundreds of millions of users and improving key business metrics, demonstrates its profound practical value and industrial impact.

\balance
\bibliographystyle{ACM-Reference-Format}
{\bibliography{sample-base}}


\begin{thebibliography}{66}


\ifx \showCODEN    \undefined \def \showCODEN     #1{\unskip}     \fi
\ifx \showISBNx    \undefined \def \showISBNx     #1{\unskip}     \fi
\ifx \showISBNxiii \undefined \def \showISBNxiii  #1{\unskip}     \fi
\ifx \showISSN     \undefined \def \showISSN      #1{\unskip}     \fi
\ifx \showLCCN     \undefined \def \showLCCN      #1{\unskip}     \fi
\ifx \shownote     \undefined \def \shownote      #1{#1}          \fi
\ifx \showarticletitle \undefined \def \showarticletitle #1{#1}   \fi
\ifx \showURL      \undefined \def \showURL       {\relax}        \fi
\providecommand\bibfield[2]{#2}
\providecommand\bibinfo[2]{#2}
\providecommand\natexlab[1]{#1}
\providecommand\showeprint[2][]{arXiv:#2}

\bibitem[Allan et~al\mbox{.}(2020)]%
        {allan2020propensity}
\bibfield{author}{\bibinfo{person}{Victoria Allan}, \bibinfo{person}{Sreeram~V Ramagopalan}, \bibinfo{person}{Jack Mardekian}, \bibinfo{person}{Aaron Jenkins}, \bibinfo{person}{Xiaoyan Li}, \bibinfo{person}{Xianying Pan}, {and} \bibinfo{person}{Xuemei Luo}.} \bibinfo{year}{2020}\natexlab{}.
\newblock \showarticletitle{Propensity score matching and inverse probability of treatment weighting to address confounding by indication in comparative effectiveness research of oral anticoagulants}.
\newblock \bibinfo{journal}{\emph{Journal of comparative effectiveness research}} \bibinfo{volume}{9}, \bibinfo{number}{9} (\bibinfo{year}{2020}), \bibinfo{pages}{603--614}.
\newblock


\bibitem[Cao et~al\mbox{.}(2024)]%
        {cao2024moment}
\bibfield{author}{\bibinfo{person}{Jiangxia Cao}, \bibinfo{person}{Shen Wang}, \bibinfo{person}{Yue Li}, \bibinfo{person}{Shenghui Wang}, \bibinfo{person}{Jian Tang}, \bibinfo{person}{Shiyao Wang}, \bibinfo{person}{Shuang Yang}, \bibinfo{person}{Zhaojie Liu}, {and} \bibinfo{person}{Guorui Zhou}.} \bibinfo{year}{2024}\natexlab{}.
\newblock \showarticletitle{Moment\&Cross: Next-Generation Real-Time Cross-Domain CTR Prediction for Live-Streaming Recommendation at Kuaishou}.
\newblock \bibinfo{journal}{\emph{arXiv preprint arXiv:2408.05709}} (\bibinfo{year}{2024}).
\newblock


\bibitem[Cao et~al\mbox{.}(2022)]%
        {cao2022samplingneedmodelinglongterm}
\bibfield{author}{\bibinfo{person}{Yue Cao}, \bibinfo{person}{XiaoJiang Zhou}, \bibinfo{person}{Jiaqi Feng}, \bibinfo{person}{Peihao Huang}, \bibinfo{person}{Yao Xiao}, \bibinfo{person}{Dayao Chen}, {and} \bibinfo{person}{Sheng Chen}.} \bibinfo{year}{2022}\natexlab{}.
\newblock \bibinfo{title}{Sampling Is All You Need on Modeling Long-Term User Behaviors for CTR Prediction}.
\newblock
\showeprint[arxiv]{2205.10249}~[cs.IR]
\urldef\tempurl%
\url{https://arxiv.org/abs/2205.10249}
\showURL{%
\tempurl}


\bibitem[Chen et~al\mbox{.}(2021a)]%
        {chen2021autodebias}
\bibfield{author}{\bibinfo{person}{Jiawei Chen}, \bibinfo{person}{Hande Dong}, \bibinfo{person}{Yang Qiu}, \bibinfo{person}{Xiangnan He}, \bibinfo{person}{Xin Xin}, \bibinfo{person}{Liang Chen}, \bibinfo{person}{Guli Lin}, {and} \bibinfo{person}{Keping Yang}.} \bibinfo{year}{2021}\natexlab{a}.
\newblock \showarticletitle{AutoDebias: Learning to debias for recommendation}. In \bibinfo{booktitle}{\emph{Proceedings of the 44th international ACM SIGIR conference on research and development in information retrieval}}. \bibinfo{pages}{21--30}.
\newblock


\bibitem[Chen et~al\mbox{.}(2023)]%
        {chen2023bias}
\bibfield{author}{\bibinfo{person}{Jiawei Chen}, \bibinfo{person}{Hande Dong}, \bibinfo{person}{Xiang Wang}, \bibinfo{person}{Fuli Feng}, \bibinfo{person}{Meng Wang}, {and} \bibinfo{person}{Xiangnan He}.} \bibinfo{year}{2023}\natexlab{}.
\newblock \showarticletitle{Bias and debias in recommender system: A survey and future directions}.
\newblock \bibinfo{journal}{\emph{ACM Transactions on Information Systems}} \bibinfo{volume}{41}, \bibinfo{number}{3} (\bibinfo{year}{2023}), \bibinfo{pages}{1--39}.
\newblock


\bibitem[Chen et~al\mbox{.}(2021c)]%
        {chen2021bias}
\bibfield{author}{\bibinfo{person}{Jiawei Chen}, \bibinfo{person}{Xiang Wang}, \bibinfo{person}{Fuli Feng}, {and} \bibinfo{person}{Xiangnan He}.} \bibinfo{year}{2021}\natexlab{c}.
\newblock \showarticletitle{Bias issues and solutions in recommender system: Tutorial on the RecSys 2021}. In \bibinfo{booktitle}{\emph{Proceedings of the 15th ACM conference on recommender systems}}. \bibinfo{pages}{825--827}.
\newblock


\bibitem[Chen et~al\mbox{.}(2024a)]%
        {chen2024treatment}
\bibfield{author}{\bibinfo{person}{Jiaju Chen}, \bibinfo{person}{Wang Wenjie}, \bibinfo{person}{Chongming Gao}, \bibinfo{person}{Peng Wu}, \bibinfo{person}{Jianxiong Wei}, {and} \bibinfo{person}{Qingsong Hua}.} \bibinfo{year}{2024}\natexlab{a}.
\newblock \showarticletitle{Treatment Effect Estimation for User Interest Exploration on Recommender Systems}. In \bibinfo{booktitle}{\emph{Proceedings of the 47th International ACM SIGIR Conference on Research and Development in Information Retrieval}}. \bibinfo{pages}{1861--1871}.
\newblock


\bibitem[Chen et~al\mbox{.}(2022)]%
        {chen2022best}
\bibfield{author}{\bibinfo{person}{Jeffrey~W Chen}, \bibinfo{person}{David~R Maldonado}, \bibinfo{person}{Brooke~L Kowalski}, \bibinfo{person}{Kara~B Miecznikowski}, \bibinfo{person}{Cynthia Kyin}, \bibinfo{person}{Jeffrey~A Gornbein}, {and} \bibinfo{person}{Benjamin~G Domb}.} \bibinfo{year}{2022}\natexlab{}.
\newblock \showarticletitle{Best practice guidelines for propensity score methods in medical research: consideration on theory, implementation, and reporting. A review}.
\newblock \bibinfo{journal}{\emph{Arthroscopy: The Journal of Arthroscopic \& Related Surgery}} \bibinfo{volume}{38}, \bibinfo{number}{2} (\bibinfo{year}{2022}), \bibinfo{pages}{632--642}.
\newblock


\bibitem[Chen et~al\mbox{.}(2021b)]%
        {chen2021endtoenduserbehaviorretrieval}
\bibfield{author}{\bibinfo{person}{Qiwei Chen}, \bibinfo{person}{Changhua Pei}, \bibinfo{person}{Shanshan Lv}, \bibinfo{person}{Chao Li}, \bibinfo{person}{Junfeng Ge}, {and} \bibinfo{person}{Wenwu Ou}.} \bibinfo{year}{2021}\natexlab{b}.
\newblock \bibinfo{title}{End-to-End User Behavior Retrieval in Click-Through RatePrediction Model}.
\newblock
\showeprint[arxiv]{2108.04468}~[cs.IR]
\urldef\tempurl%
\url{https://arxiv.org/abs/2108.04468}
\showURL{%
\tempurl}


\bibitem[Chen et~al\mbox{.}(2019)]%
        {chen2019behavior}
\bibfield{author}{\bibinfo{person}{Qiwei Chen}, \bibinfo{person}{Huan Zhao}, \bibinfo{person}{Wei Li}, \bibinfo{person}{Pipei Huang}, {and} \bibinfo{person}{Wenwu Ou}.} \bibinfo{year}{2019}\natexlab{}.
\newblock \showarticletitle{Behavior sequence transformer for e-commerce recommendation in alibaba}. In \bibinfo{booktitle}{\emph{Proceedings of the 1st international workshop on deep learning practice for high-dimensional sparse data}}. \bibinfo{pages}{1--4}.
\newblock


\bibitem[Chen et~al\mbox{.}(2025)]%
        {chen2025creator}
\bibfield{author}{\bibinfo{person}{Xiaoshuang Chen}, \bibinfo{person}{Yibo Wang}, \bibinfo{person}{Yao Wang}, \bibinfo{person}{Husheng Liu}, \bibinfo{person}{Kaiqiao Zhan}, \bibinfo{person}{Ben Wang}, {and} \bibinfo{person}{Kun Gai}.} \bibinfo{year}{2025}\natexlab{}.
\newblock \showarticletitle{Creator-Side Recommender System: Challenges, Designs, and Applications}. In \bibinfo{booktitle}{\emph{Companion Proceedings of the ACM on Web Conference 2025}}. \bibinfo{pages}{162--170}.
\newblock


\bibitem[Chen et~al\mbox{.}(2024b)]%
        {chen2024cache}
\bibfield{author}{\bibinfo{person}{Xiaoshuang Chen}, \bibinfo{person}{Gengrui Zhang}, \bibinfo{person}{Yao Wang}, \bibinfo{person}{Yulin Wu}, \bibinfo{person}{Shuo Su}, \bibinfo{person}{Kaiqiao Zhan}, {and} \bibinfo{person}{Ben Wang}.} \bibinfo{year}{2024}\natexlab{b}.
\newblock \showarticletitle{Cache-Aware Reinforcement Learning in Large-Scale Recommender Systems}. In \bibinfo{booktitle}{\emph{Companion Proceedings of the ACM Web Conference 2024}}. \bibinfo{pages}{284--291}.
\newblock


\bibitem[Cheng et~al\mbox{.}(2016)]%
        {cheng2016wide}
\bibfield{author}{\bibinfo{person}{Heng-Tze Cheng}, \bibinfo{person}{Levent Koc}, \bibinfo{person}{Jeremiah Harmsen}, \bibinfo{person}{Tal Shaked}, \bibinfo{person}{Tushar Chandra}, \bibinfo{person}{Hrishi Aradhye}, \bibinfo{person}{Glen Anderson}, \bibinfo{person}{Greg Corrado}, \bibinfo{person}{Wei Chai}, \bibinfo{person}{Mustafa Ispir}, {et~al\mbox{.}}} \bibinfo{year}{2016}\natexlab{}.
\newblock \showarticletitle{Wide \& deep learning for recommender systems}. In \bibinfo{booktitle}{\emph{Proceedings of the 1st workshop on deep learning for recommender systems}}. \bibinfo{pages}{7--10}.
\newblock


\bibitem[Dai et~al\mbox{.}(2022)]%
        {dai2022generalized}
\bibfield{author}{\bibinfo{person}{Quanyu Dai}, \bibinfo{person}{Haoxuan Li}, \bibinfo{person}{Peng Wu}, \bibinfo{person}{Zhenhua Dong}, \bibinfo{person}{Xiao-Hua Zhou}, \bibinfo{person}{Rui Zhang}, \bibinfo{person}{Rui Zhang}, {and} \bibinfo{person}{Jie Sun}.} \bibinfo{year}{2022}\natexlab{}.
\newblock \showarticletitle{A generalized doubly robust learning framework for debiasing post-click conversion rate prediction}. In \bibinfo{booktitle}{\emph{Proceedings of the 28th ACM SIGKDD Conference on Knowledge Discovery and Data Mining}}. \bibinfo{pages}{252--262}.
\newblock


\bibitem[Feng et~al\mbox{.}(2019)]%
        {feng2019deep}
\bibfield{author}{\bibinfo{person}{Yufei Feng}, \bibinfo{person}{Fuyu Lv}, \bibinfo{person}{Weichen Shen}, \bibinfo{person}{Menghan Wang}, \bibinfo{person}{Fei Sun}, \bibinfo{person}{Yu Zhu}, {and} \bibinfo{person}{Keping Yang}.} \bibinfo{year}{2019}\natexlab{}.
\newblock \showarticletitle{Deep session interest network for click-through rate prediction}.
\newblock \bibinfo{journal}{\emph{arXiv preprint arXiv:1905.06482}} (\bibinfo{year}{2019}).
\newblock


\bibitem[Gao et~al\mbox{.}(2023)]%
        {gao2023rec4ad}
\bibfield{author}{\bibinfo{person}{Jingyue Gao}, \bibinfo{person}{Shuguang Han}, \bibinfo{person}{Han Zhu}, \bibinfo{person}{Siran Yang}, \bibinfo{person}{Yuning Jiang}, \bibinfo{person}{Jian Xu}, {and} \bibinfo{person}{Bo Zheng}.} \bibinfo{year}{2023}\natexlab{}.
\newblock \showarticletitle{Rec4ad: A free lunch to mitigate sample selection bias for ads ctr prediction in taobao}. In \bibinfo{booktitle}{\emph{Proceedings of the 32nd ACM International Conference on Information and Knowledge Management}}. \bibinfo{pages}{4574--4580}.
\newblock


\bibitem[Guo et~al\mbox{.}(2017)]%
        {guo2017deepfm}
\bibfield{author}{\bibinfo{person}{Huifeng Guo}, \bibinfo{person}{Ruiming Tang}, \bibinfo{person}{Yunming Ye}, \bibinfo{person}{Zhenguo Li}, {and} \bibinfo{person}{Xiuqiang He}.} \bibinfo{year}{2017}\natexlab{}.
\newblock \showarticletitle{DeepFM: a factorization-machine based neural network for CTR prediction}.
\newblock \bibinfo{journal}{\emph{arXiv preprint arXiv:1703.04247}} (\bibinfo{year}{2017}).
\newblock


\bibitem[Guo et~al\mbox{.}(2021)]%
        {guo2021enhanced}
\bibfield{author}{\bibinfo{person}{Siyuan Guo}, \bibinfo{person}{Lixin Zou}, \bibinfo{person}{Yiding Liu}, \bibinfo{person}{Wenwen Ye}, \bibinfo{person}{Suqi Cheng}, \bibinfo{person}{Shuaiqiang Wang}, \bibinfo{person}{Hechang Chen}, \bibinfo{person}{Dawei Yin}, {and} \bibinfo{person}{Yi Chang}.} \bibinfo{year}{2021}\natexlab{}.
\newblock \showarticletitle{Enhanced doubly robust learning for debiasing post-click conversion rate estimation}. In \bibinfo{booktitle}{\emph{Proceedings of the 44th International ACM SIGIR Conference on Research and Development in Information Retrieval}}. \bibinfo{pages}{275--284}.
\newblock


\bibitem[Huang et~al\mbox{.}(2022)]%
        {huang2022different}
\bibfield{author}{\bibinfo{person}{Jin Huang}, \bibinfo{person}{Harrie Oosterhuis}, {and} \bibinfo{person}{Maarten De~Rijke}.} \bibinfo{year}{2022}\natexlab{}.
\newblock \showarticletitle{It is different when items are older: Debiasing recommendations when selection bias and user preferences are dynamic}. In \bibinfo{booktitle}{\emph{Proceedings of the fifteenth ACM international conference on web search and data mining}}. \bibinfo{pages}{381--389}.
\newblock


\bibitem[Jia et~al\mbox{.}(2025)]%
        {jia2025learn}
\bibfield{author}{\bibinfo{person}{Jian Jia}, \bibinfo{person}{Yipei Wang}, \bibinfo{person}{Yan Li}, \bibinfo{person}{Honggang Chen}, \bibinfo{person}{Xuehan Bai}, \bibinfo{person}{Zhaocheng Liu}, \bibinfo{person}{Jian Liang}, \bibinfo{person}{Quan Chen}, \bibinfo{person}{Han Li}, \bibinfo{person}{Peng Jiang}, {et~al\mbox{.}}} \bibinfo{year}{2025}\natexlab{}.
\newblock \showarticletitle{LEARN: Knowledge Adaptation from Large Language Model to Recommendation for Practical Industrial Application}. In \bibinfo{booktitle}{\emph{Proceedings of the AAAI Conference on Artificial Intelligence}}, Vol.~\bibinfo{volume}{39}. \bibinfo{pages}{11861--11869}.
\newblock


\bibitem[Juan et~al\mbox{.}(2016)]%
        {juan2016field}
\bibfield{author}{\bibinfo{person}{Yuchin Juan}, \bibinfo{person}{Yong Zhuang}, \bibinfo{person}{Wei-Sheng Chin}, {and} \bibinfo{person}{Chih-Jen Lin}.} \bibinfo{year}{2016}\natexlab{}.
\newblock \showarticletitle{Field-aware factorization machines for CTR prediction}. In \bibinfo{booktitle}{\emph{Proceedings of the 10th ACM conference on recommender systems}}. \bibinfo{pages}{43--50}.
\newblock


\bibitem[Katharopoulos et~al\mbox{.}(2020)]%
        {katharopoulos2020transformers}
\bibfield{author}{\bibinfo{person}{Angelos Katharopoulos}, \bibinfo{person}{Apoorv Vyas}, \bibinfo{person}{Nikolaos Pappas}, {and} \bibinfo{person}{Fran{\c{c}}ois Fleuret}.} \bibinfo{year}{2020}\natexlab{}.
\newblock \showarticletitle{Transformers are rnns: Fast autoregressive transformers with linear attention}. In \bibinfo{booktitle}{\emph{International conference on machine learning}}. PMLR, \bibinfo{pages}{5156--5165}.
\newblock


\bibitem[Li et~al\mbox{.}(2023)]%
        {li2023multiple}
\bibfield{author}{\bibinfo{person}{Haoxuan Li}, \bibinfo{person}{Quanyu Dai}, \bibinfo{person}{Yuru Li}, \bibinfo{person}{Yan Lyu}, \bibinfo{person}{Zhenhua Dong}, \bibinfo{person}{Xiao-Hua Zhou}, {and} \bibinfo{person}{Peng Wu}.} \bibinfo{year}{2023}\natexlab{}.
\newblock \showarticletitle{Multiple robust learning for recommendation}. In \bibinfo{booktitle}{\emph{Proceedings of the AAAI conference on artificial intelligence}}, Vol.~\bibinfo{volume}{37}. \bibinfo{pages}{4417--4425}.
\newblock


\bibitem[Li et~al\mbox{.}(2022a)]%
        {li2022tdr}
\bibfield{author}{\bibinfo{person}{Haoxuan Li}, \bibinfo{person}{Yan Lyu}, \bibinfo{person}{Chunyuan Zheng}, {and} \bibinfo{person}{Peng Wu}.} \bibinfo{year}{2022}\natexlab{a}.
\newblock \showarticletitle{TDR-CL: Targeted doubly robust collaborative learning for debiased recommendations}.
\newblock \bibinfo{journal}{\emph{arXiv preprint arXiv:2203.10258}} (\bibinfo{year}{2022}).
\newblock


\bibitem[Li et~al\mbox{.}(2022b)]%
        {li2022stabledr}
\bibfield{author}{\bibinfo{person}{Haoxuan Li}, \bibinfo{person}{Chunyuan Zheng}, {and} \bibinfo{person}{Peng Wu}.} \bibinfo{year}{2022}\natexlab{b}.
\newblock \showarticletitle{StableDR: Stabilized doubly robust learning for recommendation on data missing not at random}.
\newblock \bibinfo{journal}{\emph{arXiv preprint arXiv:2205.04701}} (\bibinfo{year}{2022}).
\newblock


\bibitem[Li et~al\mbox{.}(2025a)]%
        {li2025unbiased}
\bibfield{author}{\bibinfo{person}{Xuanlin Li}, \bibinfo{person}{Xiangyu Cai}, \bibinfo{person}{Hao Peng}, \bibinfo{person}{Jia Duan}, \bibinfo{person}{Wei Wang}, \bibinfo{person}{Zehua Zhang}, \bibinfo{person}{Changping Peng}, \bibinfo{person}{Zhangang Lin}, \bibinfo{person}{Ching Law}, {and} \bibinfo{person}{Jingping Shao}.} \bibinfo{year}{2025}\natexlab{a}.
\newblock \showarticletitle{An Unbiased Entire-Space Causal Framework for Click-Through Rate Estimation in Pre-Ranking}. In \bibinfo{booktitle}{\emph{Companion Proceedings of the ACM on Web Conference 2025}}. \bibinfo{pages}{306--314}.
\newblock


\bibitem[Li et~al\mbox{.}(2025b)]%
        {li2025survey}
\bibfield{author}{\bibinfo{person}{Yongkang Li}, \bibinfo{person}{Xingyu Zhu}, \bibinfo{person}{Yuheng Wu}, \bibinfo{person}{Wenxu Zhao}, {and} \bibinfo{person}{Xiaona Xia}.} \bibinfo{year}{2025}\natexlab{b}.
\newblock \showarticletitle{A Survey on Causal Inference-Driven Data Bias Optimization in Recommendation Systems: Principles, Opportunities and Challenges}.
\newblock \bibinfo{journal}{\emph{Wiley Interdisciplinary Reviews: Data Mining and Knowledge Discovery}} \bibinfo{volume}{15}, \bibinfo{number}{2} (\bibinfo{year}{2025}), \bibinfo{pages}{e70020}.
\newblock


\bibitem[Lin et~al\mbox{.}(2024)]%
        {lin2024mitigating}
\bibfield{author}{\bibinfo{person}{Jiaye Lin}, \bibinfo{person}{Qing Li}, \bibinfo{person}{Guorui Xie}, \bibinfo{person}{Zhongxu Guan}, \bibinfo{person}{Yong Jiang}, \bibinfo{person}{Ting Xu}, \bibinfo{person}{Zhong Zhang}, {and} \bibinfo{person}{Peilin Zhao}.} \bibinfo{year}{2024}\natexlab{}.
\newblock \showarticletitle{Mitigating Sample Selection Bias with Robust Domain Adaption in Multimedia Recommendation}. In \bibinfo{booktitle}{\emph{Proceedings of the 32nd ACM International Conference on Multimedia}}. \bibinfo{pages}{7581--7590}.
\newblock


\bibitem[Liu et~al\mbox{.}(2022)]%
        {liu2022rating}
\bibfield{author}{\bibinfo{person}{Haochen Liu}, \bibinfo{person}{Da Tang}, \bibinfo{person}{Ji Yang}, \bibinfo{person}{Xiangyu Zhao}, \bibinfo{person}{Hui Liu}, \bibinfo{person}{Jiliang Tang}, {and} \bibinfo{person}{Youlong Cheng}.} \bibinfo{year}{2022}\natexlab{}.
\newblock \showarticletitle{Rating distribution calibration for selection bias mitigation in recommendations}. In \bibinfo{booktitle}{\emph{Proceedings of the ACM web conference 2022}}. \bibinfo{pages}{2048--2057}.
\newblock


\bibitem[Liu et~al\mbox{.}(2024)]%
        {liu2024recflow}
\bibfield{author}{\bibinfo{person}{Qi Liu}, \bibinfo{person}{Kai Zheng}, \bibinfo{person}{Rui Huang}, \bibinfo{person}{Wuchao Li}, \bibinfo{person}{Kuo Cai}, \bibinfo{person}{Yuan Chai}, \bibinfo{person}{Yanan Niu}, \bibinfo{person}{Yiqun Hui}, \bibinfo{person}{Bing Han}, \bibinfo{person}{Na Mou}, {et~al\mbox{.}}} \bibinfo{year}{2024}\natexlab{}.
\newblock \showarticletitle{Recflow: An industrial full flow recommendation dataset}.
\newblock \bibinfo{journal}{\emph{arXiv preprint arXiv:2410.20868}} (\bibinfo{year}{2024}).
\newblock


\bibitem[Ma and Yu(2023)]%
        {ma2023selection}
\bibfield{author}{\bibinfo{person}{Teng Ma} {and} \bibinfo{person}{Su Yu}.} \bibinfo{year}{2023}\natexlab{}.
\newblock \showarticletitle{De-Selection Bias Recommendation Algorithm Based on Propensity Score Estimation}.
\newblock \bibinfo{journal}{\emph{Applied Sciences}} \bibinfo{volume}{13}, \bibinfo{number}{14} (\bibinfo{year}{2023}), \bibinfo{pages}{8038}.
\newblock


\bibitem[McMahan et~al\mbox{.}(2013)]%
        {mcmahan2013ad}
\bibfield{author}{\bibinfo{person}{H~Brendan McMahan}, \bibinfo{person}{Gary Holt}, \bibinfo{person}{David Sculley}, \bibinfo{person}{Michael Young}, \bibinfo{person}{Dietmar Ebner}, \bibinfo{person}{Julian Grady}, \bibinfo{person}{Lan Nie}, \bibinfo{person}{Todd Phillips}, \bibinfo{person}{Eugene Davydov}, \bibinfo{person}{Daniel Golovin}, {et~al\mbox{.}}} \bibinfo{year}{2013}\natexlab{}.
\newblock \showarticletitle{Ad click prediction: a view from the trenches}. In \bibinfo{booktitle}{\emph{Proceedings of the 19th ACM SIGKDD international conference on Knowledge discovery and data mining}}. \bibinfo{pages}{1222--1230}.
\newblock


\bibitem[Ovaisi et~al\mbox{.}(2020)]%
        {ovaisi2020correcting}
\bibfield{author}{\bibinfo{person}{Zohreh Ovaisi}, \bibinfo{person}{Ragib Ahsan}, \bibinfo{person}{Yifan Zhang}, \bibinfo{person}{Kathryn Vasilaky}, {and} \bibinfo{person}{Elena Zheleva}.} \bibinfo{year}{2020}\natexlab{}.
\newblock \showarticletitle{Correcting for selection bias in learning-to-rank systems}. In \bibinfo{booktitle}{\emph{Proceedings of the web conference 2020}}. \bibinfo{pages}{1863--1873}.
\newblock


\bibitem[Pan et~al\mbox{.}(2021)]%
        {pan2021correcting}
\bibfield{author}{\bibinfo{person}{Weishen Pan}, \bibinfo{person}{Sen Cui}, \bibinfo{person}{Hongyi Wen}, \bibinfo{person}{Kun Chen}, \bibinfo{person}{Changshui Zhang}, {and} \bibinfo{person}{Fei Wang}.} \bibinfo{year}{2021}\natexlab{}.
\newblock \showarticletitle{Correcting the user feedback-loop bias for recommendation systems}.
\newblock \bibinfo{journal}{\emph{arXiv preprint arXiv:2109.06037}} (\bibinfo{year}{2021}).
\newblock


\bibitem[Pi et~al\mbox{.}(2019)]%
        {Pi_2019}
\bibfield{author}{\bibinfo{person}{Qi Pi}, \bibinfo{person}{Weijie Bian}, \bibinfo{person}{Guorui Zhou}, \bibinfo{person}{Xiaoqiang Zhu}, {and} \bibinfo{person}{Kun Gai}.} \bibinfo{year}{2019}\natexlab{}.
\newblock \showarticletitle{Practice on Long Sequential User Behavior Modeling for Click-Through Rate Prediction}. In \bibinfo{booktitle}{\emph{Proceedings of the 25th ACM SIGKDD International Conference on Knowledge Discovery \&; Data Mining}} \emph{(\bibinfo{series}{KDD ’19})}. \bibinfo{publisher}{ACM}, \bibinfo{pages}{2671–2679}.
\newblock
\href{https://doi.org/10.1145/3292500.3330666}{doi:\nolinkurl{10.1145/3292500.3330666}}


\bibitem[Qi et~al\mbox{.}(2020)]%
        {qi2020searchbasedusermodelinglifelong}
\bibfield{author}{\bibinfo{person}{Pi Qi}, \bibinfo{person}{Xiaoqiang Zhu}, \bibinfo{person}{Guorui Zhou}, \bibinfo{person}{Yujing Zhang}, \bibinfo{person}{Zhe Wang}, \bibinfo{person}{Lejian Ren}, \bibinfo{person}{Ying Fan}, {and} \bibinfo{person}{Kun Gai}.} \bibinfo{year}{2020}\natexlab{}.
\newblock \bibinfo{title}{Search-based User Interest Modeling with Lifelong Sequential Behavior Data for Click-Through Rate Prediction}.
\newblock
\showeprint[arxiv]{2006.05639}~[cs.IR]
\urldef\tempurl%
\url{https://arxiv.org/abs/2006.05639}
\showURL{%
\tempurl}


\bibitem[Qin et~al\mbox{.}(2023)]%
        {qin2023learning}
\bibfield{author}{\bibinfo{person}{Jiarui Qin}, \bibinfo{person}{Weinan Zhang}, \bibinfo{person}{Rong Su}, \bibinfo{person}{Zhirong Liu}, \bibinfo{person}{Weiwen Liu}, \bibinfo{person}{Guangpeng Zhao}, \bibinfo{person}{Hao Li}, \bibinfo{person}{Ruiming Tang}, \bibinfo{person}{Xiuqiang He}, {and} \bibinfo{person}{Yong Yu}.} \bibinfo{year}{2023}\natexlab{}.
\newblock \showarticletitle{Learning to retrieve user behaviors for click-through rate estimation}.
\newblock \bibinfo{journal}{\emph{ACM Transactions on Information Systems}} \bibinfo{volume}{41}, \bibinfo{number}{4} (\bibinfo{year}{2023}), \bibinfo{pages}{1--31}.
\newblock


\bibitem[Qin et~al\mbox{.}(2020)]%
        {Qin_2020}
\bibfield{author}{\bibinfo{person}{Jiarui Qin}, \bibinfo{person}{Weinan Zhang}, \bibinfo{person}{Xin Wu}, \bibinfo{person}{Jiarui Jin}, \bibinfo{person}{Yuchen Fang}, {and} \bibinfo{person}{Yong Yu}.} \bibinfo{year}{2020}\natexlab{}.
\newblock \showarticletitle{User Behavior Retrieval for Click-Through Rate Prediction}. In \bibinfo{booktitle}{\emph{Proceedings of the 43rd International ACM SIGIR Conference on Research and Development in Information Retrieval}} \emph{(\bibinfo{series}{SIGIR ’20})}. \bibinfo{publisher}{ACM}, \bibinfo{pages}{2347–2356}.
\newblock
\href{https://doi.org/10.1145/3397271.3401440}{doi:\nolinkurl{10.1145/3397271.3401440}}


\bibitem[Qin et~al\mbox{.}(2022)]%
        {qin2022rankflow}
\bibfield{author}{\bibinfo{person}{Jiarui Qin}, \bibinfo{person}{Jiachen Zhu}, \bibinfo{person}{Bo Chen}, \bibinfo{person}{Zhirong Liu}, \bibinfo{person}{Weiwen Liu}, \bibinfo{person}{Ruiming Tang}, \bibinfo{person}{Rui Zhang}, \bibinfo{person}{Yong Yu}, {and} \bibinfo{person}{Weinan Zhang}.} \bibinfo{year}{2022}\natexlab{}.
\newblock \showarticletitle{Rankflow: Joint optimization of multi-stage cascade ranking systems as flows}. In \bibinfo{booktitle}{\emph{Proceedings of the 45th International ACM SIGIR Conference on Research and Development in Information Retrieval}}. \bibinfo{pages}{814--824}.
\newblock


\bibitem[Ren et~al\mbox{.}(2019)]%
        {Ren_2019}
\bibfield{author}{\bibinfo{person}{Kan Ren}, \bibinfo{person}{Jiarui Qin}, \bibinfo{person}{Yuchen Fang}, \bibinfo{person}{Weinan Zhang}, \bibinfo{person}{Lei Zheng}, \bibinfo{person}{Weijie Bian}, \bibinfo{person}{Guorui Zhou}, \bibinfo{person}{Jian Xu}, \bibinfo{person}{Yong Yu}, \bibinfo{person}{Xiaoqiang Zhu}, {and} \bibinfo{person}{Kun Gai}.} \bibinfo{year}{2019}\natexlab{}.
\newblock \showarticletitle{Lifelong Sequential Modeling with Personalized Memorization for User Response Prediction}. In \bibinfo{booktitle}{\emph{Proceedings of the 42nd International ACM SIGIR Conference on Research and Development in Information Retrieval}} \emph{(\bibinfo{series}{SIGIR ’19})}. \bibinfo{publisher}{ACM}, \bibinfo{pages}{565–574}.
\newblock
\href{https://doi.org/10.1145/3331184.3331230}{doi:\nolinkurl{10.1145/3331184.3331230}}


\bibitem[Rendle(2010)]%
        {rendle2010factorization}
\bibfield{author}{\bibinfo{person}{Steffen Rendle}.} \bibinfo{year}{2010}\natexlab{}.
\newblock \showarticletitle{Factorization machines}. In \bibinfo{booktitle}{\emph{2010 IEEE International conference on data mining}}. IEEE, \bibinfo{pages}{995--1000}.
\newblock


\bibitem[Saito(2020)]%
        {saito2020doubly}
\bibfield{author}{\bibinfo{person}{Yuta Saito}.} \bibinfo{year}{2020}\natexlab{}.
\newblock \showarticletitle{Doubly robust estimator for ranking metrics with post-click conversions}. In \bibinfo{booktitle}{\emph{Proceedings of the 14th ACM Conference on Recommender Systems}}. \bibinfo{pages}{92--100}.
\newblock


\bibitem[Schnabel et~al\mbox{.}(2016)]%
        {schnabel2016recommendations}
\bibfield{author}{\bibinfo{person}{Tobias Schnabel}, \bibinfo{person}{Adith Swaminathan}, \bibinfo{person}{Ashudeep Singh}, \bibinfo{person}{Navin Chandak}, {and} \bibinfo{person}{Thorsten Joachims}.} \bibinfo{year}{2016}\natexlab{}.
\newblock \showarticletitle{Recommendations as treatments: Debiasing learning and evaluation}. In \bibinfo{booktitle}{\emph{international conference on machine learning}}. PMLR, \bibinfo{pages}{1670--1679}.
\newblock


\bibitem[Si et~al\mbox{.}(2024)]%
        {si2024twin}
\bibfield{author}{\bibinfo{person}{Zihua Si}, \bibinfo{person}{Lin Guan}, \bibinfo{person}{ZhongXiang Sun}, \bibinfo{person}{Xiaoxue Zang}, \bibinfo{person}{Jing Lu}, \bibinfo{person}{Yiqun Hui}, \bibinfo{person}{Xingchao Cao}, \bibinfo{person}{Zeyu Yang}, \bibinfo{person}{Yichen Zheng}, \bibinfo{person}{Dewei Leng}, {et~al\mbox{.}}} \bibinfo{year}{2024}\natexlab{}.
\newblock \showarticletitle{Twin v2: Scaling ultra-long user behavior sequence modeling for enhanced ctr prediction at kuaishou}. In \bibinfo{booktitle}{\emph{Proceedings of the 33rd ACM International Conference on Information and Knowledge Management}}. \bibinfo{pages}{4890--4897}.
\newblock


\bibitem[Song et~al\mbox{.}(2023)]%
        {song2023cdr}
\bibfield{author}{\bibinfo{person}{Zijie Song}, \bibinfo{person}{Jiawei Chen}, \bibinfo{person}{Sheng Zhou}, \bibinfo{person}{Qihao Shi}, \bibinfo{person}{Yan Feng}, \bibinfo{person}{Chun Chen}, {and} \bibinfo{person}{Can Wang}.} \bibinfo{year}{2023}\natexlab{}.
\newblock \showarticletitle{CDR: Conservative doubly robust learning for debiased recommendation}. In \bibinfo{booktitle}{\emph{Proceedings of the 32nd ACM international conference on information and knowledge management}}. \bibinfo{pages}{2321--2330}.
\newblock


\bibitem[Su et~al\mbox{.}(2024b)]%
        {su2024ddpo}
\bibfield{author}{\bibinfo{person}{Hongzu Su}, \bibinfo{person}{Lichao Meng}, \bibinfo{person}{Lei Zhu}, \bibinfo{person}{Ke Lu}, {and} \bibinfo{person}{Jingjing Li}.} \bibinfo{year}{2024}\natexlab{b}.
\newblock \showarticletitle{DDPO: Direct dual propensity optimization for post-click conversion rate estimation}. In \bibinfo{booktitle}{\emph{Proceedings of the 47th International ACM SIGIR Conference on Research and Development in Information Retrieval}}. \bibinfo{pages}{1179--1188}.
\newblock


\bibitem[Su et~al\mbox{.}(2024a)]%
        {su2024rpaf}
\bibfield{author}{\bibinfo{person}{Shuo Su}, \bibinfo{person}{Xiaoshuang Chen}, \bibinfo{person}{Yao Wang}, \bibinfo{person}{Yulin Wu}, \bibinfo{person}{Ziqiang Zhang}, \bibinfo{person}{Kaiqiao Zhan}, \bibinfo{person}{Ben Wang}, {and} \bibinfo{person}{Kun Gai}.} \bibinfo{year}{2024}\natexlab{a}.
\newblock \showarticletitle{RPAF: A Reinforcement Prediction-Allocation Framework for Cache Allocation in Large-Scale Recommender Systems}. In \bibinfo{booktitle}{\emph{Proceedings of the 18th ACM Conference on Recommender Systems}}. \bibinfo{pages}{670--679}.
\newblock


\bibitem[Wang et~al\mbox{.}(2025)]%
        {wang2025findrec}
\bibfield{author}{\bibinfo{person}{Maolin Wang}, \bibinfo{person}{Yutian Xiao}, \bibinfo{person}{Binhao Wang}, \bibinfo{person}{Sheng Zhang}, \bibinfo{person}{Shanshan Ye}, \bibinfo{person}{Wanyu Wang}, \bibinfo{person}{Hongzhi Yin}, \bibinfo{person}{Ruocheng Guo}, {and} \bibinfo{person}{Zenglin Xu}.} \bibinfo{year}{2025}\natexlab{}.
\newblock \showarticletitle{FindRec: Stein-Guided Entropic Flow for Multi-Modal Sequential Recommendation}. In \bibinfo{booktitle}{\emph{Proceedings of the 31st ACM SIGKDD Conference on Knowledge Discovery and Data Mining V. 2}}. \bibinfo{pages}{3008--3018}.
\newblock


\bibitem[Wang et~al\mbox{.}(2017)]%
        {wang2017deep}
\bibfield{author}{\bibinfo{person}{Ruoxi Wang}, \bibinfo{person}{Bin Fu}, \bibinfo{person}{Gang Fu}, {and} \bibinfo{person}{Mingliang Wang}.} \bibinfo{year}{2017}\natexlab{}.
\newblock \showarticletitle{Deep \& cross network for ad click predictions}.
\newblock In \bibinfo{booktitle}{\emph{Proceedings of the ADKDD'17}}. \bibinfo{pages}{1--7}.
\newblock


\bibitem[Wang et~al\mbox{.}(2021a)]%
        {wang2021dcn}
\bibfield{author}{\bibinfo{person}{Ruoxi Wang}, \bibinfo{person}{Rakesh Shivanna}, \bibinfo{person}{Derek Cheng}, \bibinfo{person}{Sagar Jain}, \bibinfo{person}{Dong Lin}, \bibinfo{person}{Lichan Hong}, {and} \bibinfo{person}{Ed Chi}.} \bibinfo{year}{2021}\natexlab{a}.
\newblock \showarticletitle{Dcn v2: Improved deep \& cross network and practical lessons for web-scale learning to rank systems}. In \bibinfo{booktitle}{\emph{Proceedings of the web conference 2021}}. \bibinfo{pages}{1785--1797}.
\newblock


\bibitem[Wang et~al\mbox{.}(2019)]%
        {wang2019doubly}
\bibfield{author}{\bibinfo{person}{Xiaojie Wang}, \bibinfo{person}{Rui Zhang}, \bibinfo{person}{Yu Sun}, {and} \bibinfo{person}{Jianzhong Qi}.} \bibinfo{year}{2019}\natexlab{}.
\newblock \showarticletitle{Doubly robust joint learning for recommendation on data missing not at random}. In \bibinfo{booktitle}{\emph{International Conference on Machine Learning}}. PMLR, \bibinfo{pages}{6638--6647}.
\newblock


\bibitem[Wang et~al\mbox{.}(2021b)]%
        {wang2021combating}
\bibfield{author}{\bibinfo{person}{Xiaojie Wang}, \bibinfo{person}{Rui Zhang}, \bibinfo{person}{Yu Sun}, {and} \bibinfo{person}{Jianzhong Qi}.} \bibinfo{year}{2021}\natexlab{b}.
\newblock \showarticletitle{Combating selection biases in recommender systems with a few unbiased ratings}. In \bibinfo{booktitle}{\emph{Proceedings of the 14th ACM International Conference on Web Search and Data Mining}}. \bibinfo{pages}{427--435}.
\newblock


\bibitem[Wei et~al\mbox{.}(2024)]%
        {wei2024enhancing}
\bibfield{author}{\bibinfo{person}{Jianping Wei}, \bibinfo{person}{Yujie Zhou}, \bibinfo{person}{Zhengwei Wu}, {and} \bibinfo{person}{Ziqi Liu}.} \bibinfo{year}{2024}\natexlab{}.
\newblock \showarticletitle{Enhancing pre-ranking performance: Tackling intermediary challenges in multi-stage cascading recommendation systems}. In \bibinfo{booktitle}{\emph{Proceedings of the 30th ACM SIGKDD Conference on Knowledge Discovery and Data Mining}}. \bibinfo{pages}{5950--5958}.
\newblock


\bibitem[Xiao et~al\mbox{.}(2025)]%
        {xiao2025marsmodalityalignedretrievalsequence}
\bibfield{author}{\bibinfo{person}{Yutian Xiao}, \bibinfo{person}{Shukuan Wang}, \bibinfo{person}{Binhao Wang}, \bibinfo{person}{Zhao Zhang}, \bibinfo{person}{Yanze Zhang}, \bibinfo{person}{Shanqi Liu}, \bibinfo{person}{Chao Feng}, \bibinfo{person}{Xiang Li}, {and} \bibinfo{person}{Fuzhen Zhuang}.} \bibinfo{year}{2025}\natexlab{}.
\newblock \bibinfo{title}{MARS: Modality-Aligned Retrieval for Sequence Augmented CTR Prediction}.
\newblock
\showeprint[arxiv]{2509.01184}~[cs.IR]
\urldef\tempurl%
\url{https://arxiv.org/abs/2509.01184}
\showURL{%
\tempurl}


\bibitem[Xu et~al\mbox{.}(2022b)]%
        {xu2022dually}
\bibfield{author}{\bibinfo{person}{Chen Xu}, \bibinfo{person}{Jun Xu}, \bibinfo{person}{Xu Chen}, \bibinfo{person}{Zhenghua Dong}, {and} \bibinfo{person}{Ji-Rong Wen}.} \bibinfo{year}{2022}\natexlab{b}.
\newblock \showarticletitle{Dually enhanced propensity score estimation in sequential recommendation}. In \bibinfo{booktitle}{\emph{Proceedings of the 31st ACM International Conference on Information \& Knowledge Management}}. \bibinfo{pages}{2260--2269}.
\newblock


\bibitem[Xu et~al\mbox{.}(2022a)]%
        {xu2022ukd}
\bibfield{author}{\bibinfo{person}{Zixuan Xu}, \bibinfo{person}{Penghui Wei}, \bibinfo{person}{Weimin Zhang}, \bibinfo{person}{Shaoguo Liu}, \bibinfo{person}{Liang Wang}, {and} \bibinfo{person}{Bo Zheng}.} \bibinfo{year}{2022}\natexlab{a}.
\newblock \showarticletitle{Ukd: Debiasing conversion rate estimation via uncertainty-regularized knowledge distillation}. In \bibinfo{booktitle}{\emph{Proceedings of the ACM Web Conference 2022}}. \bibinfo{pages}{2078--2087}.
\newblock


\bibitem[Yu et~al\mbox{.}(2021)]%
        {yu2021self}
\bibfield{author}{\bibinfo{person}{Wenhui Yu}, \bibinfo{person}{Xiao Lin}, \bibinfo{person}{Jinfei Liu}, \bibinfo{person}{Junfeng Ge}, \bibinfo{person}{Wenwu Ou}, {and} \bibinfo{person}{Zheng Qin}.} \bibinfo{year}{2021}\natexlab{}.
\newblock \showarticletitle{Self-propagation graph neural network for recommendation}.
\newblock \bibinfo{journal}{\emph{IEEE Transactions on Knowledge and Data Engineering}} \bibinfo{volume}{34}, \bibinfo{number}{12} (\bibinfo{year}{2021}), \bibinfo{pages}{5993--6002}.
\newblock


\bibitem[Zhang et~al\mbox{.}(2025)]%
        {zhang2025mitigating}
\bibfield{author}{\bibinfo{person}{Fan Zhang}, \bibinfo{person}{Wenjie Luo}, {and} \bibinfo{person}{Xiudan Yang}.} \bibinfo{year}{2025}\natexlab{}.
\newblock \showarticletitle{Mitigating Selection Bias in Recommendation Systems Through Sentiment Analysis and Dynamic Debiasing}.
\newblock \bibinfo{journal}{\emph{Applied Sciences}} \bibinfo{volume}{15}, \bibinfo{number}{8} (\bibinfo{year}{2025}), \bibinfo{pages}{4170}.
\newblock


\bibitem[Zhang et~al\mbox{.}(2020)]%
        {zhang2020large}
\bibfield{author}{\bibinfo{person}{Wenhao Zhang}, \bibinfo{person}{Wentian Bao}, \bibinfo{person}{Xiao-Yang Liu}, \bibinfo{person}{Keping Yang}, \bibinfo{person}{Quan Lin}, \bibinfo{person}{Hong Wen}, {and} \bibinfo{person}{Ramin Ramezani}.} \bibinfo{year}{2020}\natexlab{}.
\newblock \showarticletitle{Large-scale causal approaches to debiasing post-click conversion rate estimation with multi-task learning}. In \bibinfo{booktitle}{\emph{Proceedings of the web conference 2020}}. \bibinfo{pages}{2775--2781}.
\newblock


\bibitem[Zhao et~al\mbox{.}(2025)]%
        {zhao2025hybrid}
\bibfield{author}{\bibinfo{person}{Binglei Zhao}, \bibinfo{person}{Houying Qi}, \bibinfo{person}{Guang Xu}, \bibinfo{person}{Mian Ma}, \bibinfo{person}{Xiwei Zhao}, \bibinfo{person}{Feng Mei}, \bibinfo{person}{Sulong Xu}, {and} \bibinfo{person}{Jinghe Hu}.} \bibinfo{year}{2025}\natexlab{}.
\newblock \showarticletitle{A Hybrid Cross-Stage Coordination Pre-ranking Model for Online Recommendation Systems}. In \bibinfo{booktitle}{\emph{Companion Proceedings of the ACM on Web Conference 2025}}. \bibinfo{pages}{621--630}.
\newblock


\bibitem[Zheng et~al\mbox{.}(2024)]%
        {zheng2024full}
\bibfield{author}{\bibinfo{person}{Kai Zheng}, \bibinfo{person}{Haijun Zhao}, \bibinfo{person}{Rui Huang}, \bibinfo{person}{Beichuan Zhang}, \bibinfo{person}{Na Mou}, \bibinfo{person}{Yanan Niu}, \bibinfo{person}{Yang Song}, \bibinfo{person}{Hongning Wang}, {and} \bibinfo{person}{Kun Gai}.} \bibinfo{year}{2024}\natexlab{}.
\newblock \showarticletitle{Full stage learning to rank: A unified framework for multi-stage systems}. In \bibinfo{booktitle}{\emph{Proceedings of the ACM Web Conference 2024}}. \bibinfo{pages}{3621--3631}.
\newblock


\bibitem[Zhou et~al\mbox{.}(2018a)]%
        {zhou2018deepevolutionnetworkclickthrough}
\bibfield{author}{\bibinfo{person}{Guorui Zhou}, \bibinfo{person}{Na Mou}, \bibinfo{person}{Ying Fan}, \bibinfo{person}{Qi Pi}, \bibinfo{person}{Weijie Bian}, \bibinfo{person}{Chang Zhou}, \bibinfo{person}{Xiaoqiang Zhu}, {and} \bibinfo{person}{Kun Gai}.} \bibinfo{year}{2018}\natexlab{a}.
\newblock \bibinfo{title}{Deep Interest Evolution Network for Click-Through Rate Prediction}.
\newblock
\showeprint[arxiv]{1809.03672}~[stat.ML]
\urldef\tempurl%
\url{https://arxiv.org/abs/1809.03672}
\showURL{%
\tempurl}


\bibitem[Zhou et~al\mbox{.}(2019)]%
        {zhou2019deep}
\bibfield{author}{\bibinfo{person}{Guorui Zhou}, \bibinfo{person}{Na Mou}, \bibinfo{person}{Ying Fan}, \bibinfo{person}{Qi Pi}, \bibinfo{person}{Weijie Bian}, \bibinfo{person}{Chang Zhou}, \bibinfo{person}{Xiaoqiang Zhu}, {and} \bibinfo{person}{Kun Gai}.} \bibinfo{year}{2019}\natexlab{}.
\newblock \showarticletitle{Deep interest evolution network for click-through rate prediction}. In \bibinfo{booktitle}{\emph{Proceedings of the AAAI conference on artificial intelligence}}, Vol.~\bibinfo{volume}{33}. \bibinfo{pages}{5941--5948}.
\newblock


\bibitem[Zhou et~al\mbox{.}(2018b)]%
        {zhou2018deep}
\bibfield{author}{\bibinfo{person}{Guorui Zhou}, \bibinfo{person}{Xiaoqiang Zhu}, \bibinfo{person}{Chenru Song}, \bibinfo{person}{Ying Fan}, \bibinfo{person}{Han Zhu}, \bibinfo{person}{Xiao Ma}, \bibinfo{person}{Yanghui Yan}, \bibinfo{person}{Junqi Jin}, \bibinfo{person}{Han Li}, {and} \bibinfo{person}{Kun Gai}.} \bibinfo{year}{2018}\natexlab{b}.
\newblock \showarticletitle{Deep interest network for click-through rate prediction}. In \bibinfo{booktitle}{\emph{Proceedings of the 24th ACM SIGKDD international conference on knowledge discovery \& data mining}}. \bibinfo{pages}{1059--1068}.
\newblock


\bibitem[Zhu et~al\mbox{.}(2023)]%
        {zhu2023dcmt}
\bibfield{author}{\bibinfo{person}{Feng Zhu}, \bibinfo{person}{Mingjie Zhong}, \bibinfo{person}{Xinxing Yang}, \bibinfo{person}{Longfei Li}, \bibinfo{person}{Lu Yu}, \bibinfo{person}{Tiehua Zhang}, \bibinfo{person}{Jun Zhou}, \bibinfo{person}{Chaochao Chen}, \bibinfo{person}{Fei Wu}, \bibinfo{person}{Guanfeng Liu}, {et~al\mbox{.}}} \bibinfo{year}{2023}\natexlab{}.
\newblock \showarticletitle{Dcmt: a direct entire-space causal multi-task framework for post-click conversion estimation}. In \bibinfo{booktitle}{\emph{2023 IEEE 39th International Conference on Data Engineering (ICDE)}}. IEEE, \bibinfo{pages}{3113--3125}.
\newblock


\bibitem[Zou et~al\mbox{.}(2022)]%
        {zou2022approximated}
\bibfield{author}{\bibinfo{person}{Lixin Zou}, \bibinfo{person}{Changying Hao}, \bibinfo{person}{Hengyi Cai}, \bibinfo{person}{Shuaiqiang Wang}, \bibinfo{person}{Suqi Cheng}, \bibinfo{person}{Zhicong Cheng}, \bibinfo{person}{Wenwen Ye}, \bibinfo{person}{Simiu Gu}, {and} \bibinfo{person}{Dawei Yin}.} \bibinfo{year}{2022}\natexlab{}.
\newblock \showarticletitle{Approximated doubly robust search relevance estimation}. In \bibinfo{booktitle}{\emph{Proceedings of the 31st ACM International Conference on Information \& Knowledge Management}}. \bibinfo{pages}{3756--3765}.
\newblock


\end{thebibliography}
\end{document}